%% file: Jalota.ea.EAAMO_22.tex
\documentclass[11pt]{article}
\usepackage{amsmath,inputenc}

\title{Matching with Transfers under Distributional Constraints}
\author{Devansh Jalota$^\dagger$, Michael Ostrovsky$^\ddagger$, and Marco Pavone$^*$
\thanks{$^\dagger$ Institute for Computational and Mathematical Engineering, Stanford University; {\tt djalota@stanford.edu}.}%
\thanks{$^\ddagger$ Graduate School of Business, Stanford University; {\tt ost@stanford.edu}.}%
\thanks{$^*$ Department of Aeronautics and Astronautics, Stanford University; {\tt pavone@stanford.edu}.}
}

\date{April 2022}

\input{tex/preamble}
\input{tex/macros}

\begin{document}

\maketitle

\begin{abstract}
    We study two-sided many-to-one matching markets with transferable utilities, e.g., labor and rental housing markets, in which money can exchange hands between agents, subject to distributional constraints on the set of feasible allocations. In such markets, we establish the efficiency of equilibrium arrangements, specified by an assignment and transfers between agents on the two sides of the market, and study the conditions on the distributional constraints and agent preferences under which equilibria exist and can be computed efficiently. To this end, we first consider the setting when the number of institutions (e.g., firms in a labor market) is one and show that equilibrium arrangements exist irrespective of the nature of the constraint structure or the agents' preferences. However, equilibrium arrangements may not exist in markets with multiple institutions even when agents on each side have linear (or additively separable) preferences over agents on the other side. Thus, for markets with linear preferences, we study sufficient conditions on the constraint structure that guarantee the existence of equilibria using linear programming duality. Our linear programming approach not only generalizes that of Shapley and Shubik~\cite{shapley1971assignment} in the one-to-one matching setting to the many-to-one matching setting under distributional constraints but also provides a method to compute market equilibria efficiently.
\end{abstract}

\section{Introduction}

Many real-world two-sided matching markets with transferable utilities, wherein money can exchange hands between agents, are subject to distributional constraints, e.g., quotas on the number of individuals who can be assigned to given institutions. In labor markets, firms are often constrained with regards to the types of workers that they can hire, e.g., in the United States, the government has designed affirmative action policies to ``increase the representation of minorities'' and people with disabilities in the labor force~\cite{AA-Weber}. Similar quotas on hiring have been imposed in various states across India that give ``special provisions for employment based on language''~\cite{jobReservationIndia}, and ensure enough employment opportunities for locals~\cite{jobReservationHaryana}. On the other hand, in rental housing markets, constraints on the type of people that homeowners can lease homes to arise through affordable housing programs, which mandate that homeowners reserve a certain proportion of houses for people with low incomes~\cite{housing-markets}.

In such matching markets with distributional constraints on the set of feasible allocations, traditional theory on two-sided matching under transferable utilities falls short. As a result, in this work, we study such matching markets by investigating the conditions on the constraint structure and agents' preferences under which equilibria, which we henceforth refer to as \emph{stable arrangements}\footnote{A stable arrangement is specified by an assignment and transfers between agents in the market. The specific stability concept that we are concerned with is elaborated on in Section~\ref{sec:defs}, and reduces to the standard stability concept of Kelso and Crawford~\cite{kelso1982job} if there are no distributional constraints.}, exist and can be computed efficiently. In particular, we provide a method to efficiently compute stable arrangements by establishing a connection between linear programming duality and stability in such two-sided matching markets and also study the existence of stable arrangements in settings when the \emph{substitutes condition}\footnote{The substitutes condition is often a critical condition for the existence of stable arrangements~\cite{kelso1982job,hatfield2005matching}, and we refer to Section~\ref{subsec:vioSubsttutes} for a definition of this term in the context of two-sided matching markets under distributional constraints.} may not hold. We further show that stable arrangements correspond to efficient assignments between agents and institutions, i.e., they maximize the cumulative payoff received by all agents in the market. We note here that as in~\cite{taxation-matching} we present our model and analysis in the language of labor markets, wherein firms on one side of the market pay salaries (transfers) to workers on the other side. However, we mention that our results may also be relevant for other two-sided many-to-one matching markets.

\subsection{Our Contributions}

In this work, we study a two-sided many-to-one matching market wherein agents (firms and workers) have heterogeneous preferences, constraints are imposed on the set of feasible allocations, and transfers are allowed between agents on the two sides of the market. In such markets, we investigate the formation of stable arrangements under two notions of stability, one where we require the individual rationality of firms, which is our working definition throughout the paper, and another where we relax this condition (see Section~\ref{sec:lowerBoundGeneral}).

We begin our study of stable arrangements in such matching markets by establishing a connection between stability and the efficiency of the resulting assignment of workers to firms. In particular, we show that an assignment corresponding to any stable arrangement achieves the highest cumulative payoff for all firms and workers combined.

We then investigate the conditions on the distributional constraints and agent preferences under which stable arrangements exist and can be computed efficiently. To this end, we first consider the setting of one firm and show that stable arrangements exist irrespective of the nature of agents’ preferences or the constraint structure. We mention that this result holds despite the violation of the substitutes condition (see Section~\ref{subsec:vioSubsttutes}). However, we then show that this general result on the existence of stable arrangements does not extend to the setting of multiple firms, even when firms have linear (or additively separable) preferences over workers.

As a result, in the multiple firm setting, we focus on markets where firms have linear preferences over workers and provide sufficient conditions on the constraint structure for stable arrangements to exist. In this setting, we develop a linear programming approach to compute stable arrangements in polynomial time in the number of firms, workers, and the cardinality of the constraint set. Our approach provides a connection between linear programming duality and the existence of stable arrangements, which generalizes the duality theory approach of Shapley and Shubik~\cite{shapley1971assignment} in the one-to-one matching setting to the many-to-one matching setting under distributional constraints. 

\paragraph{Organization}

This paper is organized as follows. Section~\ref{sec:litreview} reviews related literature. We then present our matching market model, introduce the notions of stability and efficiency, and establish the efficiency of stable arrangements in Section~\ref{sec:modelMain}. Then, in Section~\ref{sec:oneFirmExistence}, we consider the setting of one firm and show that stable arrangements exist under general feasibility constraints and agent preferences. Next, we consider the setting of multiple firms in Section~\ref{sec:multiFirms}, where we establish a connection between linear programming duality and the existence of stable arrangements. Finally, we conclude the paper and provide directions for future work in Section~\ref{sec:conclusion}.

\section{Related Literature} \label{sec:litreview}

Two-sided matching markets have been a cornerstone for market design ever since the seminal work of Gale and Shapley~\cite{gale1962college}, whose work has influenced the matching of individuals to institutions across various markets, e.g., assigning students to schools~\cite{roth2008deferred,abdulkadiroglu2006changing} or doctors to hospitals~\cite{roth1986allocation,al-residency-matching}. While Gale and Shapley~\cite{gale1962college} considered a pure matching problem wherein agents cannot use money as a means to compensate other agents in the market, Shapley and Shubik~\cite{shapley1971assignment} initiated a parallel line of research on two-sided matching with transferable utilities. In their model, money enters each agent's utility function linearly, and they use duality theory to show the existence of stable arrangements. Building on Shapley and Shubik's work~\cite{shapley1971assignment}, Crawford and Knoer~\cite{crawford1981job}, in the context of a labor market, developed a salary adjustment process akin to Gale and Shapley's~\cite{gale1962college} deferred acceptance algorithm. Later, Kelso and Crawford~\cite{kelso1982job} considered a more general class of economies with heterogeneous firms and workers and showed that this salary adjustment process converges to the equilibrium under a substitutes condition on the preferences of the firms. 

As with the above works, we also study two-sided matching markets with transferable utilities; however, we additionally consider distributional constraints on the set of feasible allocations. In particular, as opposed to Kelso and Crawford~\cite{kelso1982job}, we study the existence of stable arrangements under agent preferences and constraints that do not necessarily satisfy the substitutes condition. Furthermore, we extend the duality theory approach of Shapley and Shubik~\cite{shapley1971assignment} in the one-to-one matching setting to compute stable arrangements in the many-to-one matching setting under distributional constraints.

Most existing works on matching under distributional constraints have focused on the non-transferable utility setting with applications in school choice~\cite{abdulkadirouglu2005college,kojima2012school,ehlers2014school,hybm-2011}, refugee resettlement~\cite{delacretaz2016refugee,delacretaz2019matching}, and the assignment of army cadets to army branches~\cite{sonmez2013bidding,sonmez2013matching}. In such matching markets, \emph{stable matchings}\footnote{Under non-transferable utilities, we refer to equilibria as stable matchings, rather than stable arrangements, since no transfers are involved.} may not exist, and determining whether one does is NP-complete~\cite{BIRO20103136,approx-stable-general} for general constraint structures. As a result, several works have focused on investigating the set of constraints under which desirable mechanisms, e.g., variants of the deferred acceptance algorithm that improve upon the efficiency, stability, and incentive properties of existing matching algorithms under distributional constraints~\cite{kamada2017stability,kamada2015efficient,hard-distr-constraints,approx-stable-general}, can be designed. For instance, Kamada and Kojima~\cite{kamada2018stability} show that there exists a mechanism that is both stable and strategy-proof for agents on one side of the market if the constraints are \emph{hierarchical}. Similarly, Bing et al.~\cite{bilemi-2004} and Gul et al.~\cite{gul2019market} show that imposing floor and ceiling constraints on members of a hierarchical cover preserves the substitutes condition. In line with the above works on two-sided matching markets under distributional constraints, we also study the conditions on the constraint structure under with stable arrangements exist and can be computed efficiently. However, we reiterate that, unlike the above works that assume non-transferable utilities, we consider the transferable utility setting~\cite{shapley1971assignment,kelso1982job,crawford1981job,koopmans1957assignment}. 

In the transferable utility setting, Kojima et al.~\cite{jobMatchingConstraints2020} study two-sided matching markets subject to distributional constraints. In particular, Kojima et al.~\cite{jobMatchingConstraints2020} characterize the necessary and sufficient conditions on the constraint structure under which the substitutes condition holds, which guarantees the existence of stable arrangements. As in Kojima et al.~\cite{jobMatchingConstraints2020}, we also consider two-sided matching markets with transferable utilities that are subject to distributional constraints but provide an alternate perspective to and build on their work in multiple directions. First, we study the existence of stable arrangements in settings when the substitutes condition may not hold. Next, we present a method to efficiently compute stable arrangements using linear programming duality under constraint structures that are analogous to those that are necessary and sufficient for the preservation of the substitutes condition in Kojima et al.~\cite{jobMatchingConstraints2020}. This connection to linear programming not only provides a method to compute stable arrangements efficiently but also provides an alternate proof to that in Kojima et al.~\cite{jobMatchingConstraints2020} for the existence of stable arrangements in the setting when firms have linear preferences over workers. Finally, we mention that we study the existence of stable arrangements under two notions of stability, one which is considered in Kojima et al.~\cite{jobMatchingConstraints2020}, and another where we additionally require the individual rationality of firms.


\section{Model, Definitions, and Efficiency of Stable Arrangements} \label{sec:modelMain}

In this section, we present our model of two-sided many-to-one matching markets with transferable utilities and distributional constraints, introduce the notions of stability and efficiency that we use within this work, and show that assignments corresponding to stable arrangements are efficient.

\subsection{General Model} \label{subsec:generalModel}

We study a two-sided many-to-one matching market where (i) constraints are imposed on the set of feasible allocations, (ii) agents on each side of the market have heterogeneous preferences, and (iii) transfers are allowed between agents on the two sides of the market. Agents on one side of the market are referred to as firms, denoted $f \in F$, while agents on the other side are referred to as workers, denoted as $w \in W$. Each worker can be assigned (or matched) to at most one firm, while each firm can hire any feasible set of workers for that firm. For each firm $f$, we denote the feasibility collection $\T_f \subseteq 2^W$ as the collection of all subsets of workers that are feasible for that firm, i.e., any set of workers $D \in \T_f$ satisfies the distributional constraints on hiring for a firm $f$. We provide an example of such distributional constraints at the end of this section.

The agents on both sides of the market derive value from being matched to agents on the other side. To model agent preferences, we let $a_{w,f}$ be the value that worker $w$ obtains when matched to a firm $f$, and let $b_{D, f}$ be the value firm $f$ obtains when matched with a feasible set of workers $D \in \T_f$. Furthermore, without loss of generality, we normalize the value of agents that are unmatched to zero, i.e., $a_{w,\emptyset} = 0$ and $b_{\emptyset, f} = 0$, where $\emptyset$ denotes the fact that the corresponding agent is unmatched.

An assignment of workers to firms is described by a matrix $\Xb=(x_{w,f})_{w \in W, f \in F}$, where $x_{w,f} \in \{0, 1 \}$ denotes whether worker $w$ is assigned to firm $f$. Such an assignment $\Xb$ is \emph{feasible} if each worker is assigned to at most one firm and the set of workers $D$ assigned to each firm $f$ satisfies its distributional constraints, i.e., $D \in \T_f$, or firm $f$ is unmatched, i.e., $D = \emptyset$. For the ease of notation, let $f_w \in F \cup \emptyset$ be the firm that a worker $w$ is matched to (or that $w$ is unmatched) and $D_f \in \T_f \cup \emptyset$ be the set of workers that a firm $f$ is matched to (or that $f$ is unmatched) under an assignment $\Xb$.

In this market, we allow for transfers, which we refer to as salaries, between agents on the two sides of the market. We denote $s_{w,f}$ as the salary received by worker $w$ from firm $f$, and let $\s_f \in \mathbb{R}^{|W|}$ be the (prospective) salaries paid by firm $f$ to all workers irrespective of whether they are matched or not. We mention that the salaries can take on any real number and thus be negative. In particular, if the salary $s_{w,f} < 0$ for some worker $w$ and firm $f$, then we interpret this as a positive transfer received by firm $f$ from worker $w$.

Each agent's payoff is assumed to be quasi-linear in the salaries, and they are assumed to only derive value from the agent(s) they match with on the other side of the market. Under an assignment $\Xb$ and salary profile $\s_f$ for each firm $f$, the payoff $u_w$ for a worker $w$ and $v_f$ for a firm $f$ is given by
\begin{align}
    &u_w(\Xb, \{ \s_f \}_{f \in F}) = a_{w,f_w} + s_{w,f_w}, \label{eq:workerPayoff} \\ 
    &v_f(\Xb, \{ \s_f \}_{f \in F}) = b_{D_f, f} - \sum_{w \in D_f} s_{w,f}, \label{eq:firmPayoff}
\end{align}
where each worker $w$ is assigned to firm $f_w$, and a set of workers $D_f$ is assigned to each firm $f$. We note from the above payoff relation for the firms that the total salary paid by the firm to a set of workers $D_f$ is the sum of the salaries paid to each worker $w \in D_f$ by that firm. For brevity, we denote $\u \in \mathbb{R}^{|W|}$ and $\v \in \mathbb{R}^{|F|}$ as the vector of payoffs of the workers and firms, respectively. Furthermore, for notational simplicity, we will often drop the dependence of the payoffs of the workers and firms on the assignment and salary profile if it is sufficiently clear from the context.

We mention that in this more general two-sided matching setting we do not impose any structure on agents' preferences or the distributional constraints that define the feasibility collection $\T_f$ for each firm $f$. However, to elucidate our model, we present a class of agent preferences and distributional constraints, which will serve as the main focus of Section~\ref{sec:multiFirms} of this paper.

\paragraph{An Example of Agent Preferences and Distributional Constraints} We consider the setting wherein firms have linear (or additively separable) preferences over workers and the distributional constraints are described by upper and lower bound constraints on hiring. To model each firm's preferences, let $c_{w,f}$ be the value firm $f$ receives when hiring worker $w$. Then, for any feasible set of workers $D \in \T_f \cup \emptyset$, the total value received by firm $f$ is $b_{D,f} = \sum_{w \in D} c_{w,f}$, which is a sum of the values received by the firm from each individual worker in the set $D$. Furthermore, for conciseness, we denote $\alpha_{w,f} = a_{w,f} + c_{w,f}$ as the match value for the worker-firm pair $(w,f)$.

We now describe the upper and lower bound constraints on hiring that define the feasibility collection $\T_f$. To this end, for each firm $f$ let the constraint set $\H_f$ denote the collection of all subsets of workers over which the firm has hiring constraints. Furthermore, for each set of workers $D \in \H_f$ let $\underline{\lambda}_{D} \leq \Bar{\lambda}_{D}$ be non-negative integers and $\llambda_f = (\underline{\llambda}_f, \Bar{\llambda}_f) = (\underline{\lambda}_{D}, \Bar{\lambda}_{D})_{D \in \H_f}$ be the vector of hiring quotas for that firm. Then, for each firm $f$, the constraints we consider are of the form 
\begin{align*}
    \underline{\lambda}_{D} \leq \sum_{w \in {D}} x_{w,f} \leq \Bar{\lambda}_{D}, \quad \text{for all } D \in \H_f.
\end{align*}
That is, for a set of workers $D \in \H_f$ with quotas $\underline{\lambda}_{D}, \Bar{\lambda}_{D}$, the total number of workers belonging to set $D$ that are assigned to firm $f$ must lie between the floor $\underline{\lambda}_{D}$ and ceiling $\Bar{\lambda}_{D}$. In this case, the feasibility collection $\T_f$ for each firm $f$ is the collection of all subsets of workers that satisfy the upper and lower bound hiring constraints defined by the constraint structure $\H_f$ and quotas $\llambda_f$.

\subsection{Definitions} \label{sec:defs}

Given an assignment $\Xb$ and a vector $\s_f$ of salaries for each firm $f$, we define an \emph{arrangement} by a tuple $(\Xb, \{ \s_f \}_{f \in F})$, and now introduce some key definitions regarding feasibility, individual rationality, stability and efficiency, which we will leverage throughout this work.

An arrangement $(\Xb, \{ \s_f \}_{f \in F})$ is defined to be \emph{feasible} if the assignment $\Xb$ is feasible.

A feasible arrangement is \emph{individually rational} for workers (firms) if the resulting payoffs for all the workers (firms) is non-negative, i.e., $\u \geq \0$ ($\v \geq \0$). Recall that the payoffs of the workers and firms are related to the assignment $\Xb$ and the salaries $\s_f$ through Equations~\eqref{eq:workerPayoff} and~\eqref{eq:firmPayoff}, respectively.

Finally, we introduce the notion of stability we use in this work, which is analogous to that considered in Kelso and Crawford~\cite{kelso1982job}. In particular, stable arrangements are those under which no firm $f$ and \emph{feasible} group of workers $D \in \T_f \cup \emptyset$ that can form a coalition and deviate in a manner that would make all agents in the coalition better off.

\begin{definition} [Stable Arrangement] \label{def:stability}
A feasible arrangement $(\Xb, \{ \s_f \}_{f \in F})$ is stable if it is individually rational for all workers and firms, and there is no firm-worker set $\{f \} \cup D$, where $D$ is feasible for firm $f$, i.e., $D \in \T_f \cup \emptyset$, and vector of salaries $\r_f$, such that
\begin{align*}
    &a_{w,f} + r_{w,f} \geq a_{w,f_w} + s_{w,f_w}, \quad \text{for all } w \in D, \text{ and } \\
    &b_{D, f} - \sum_{w \in D} s_{w,f} \geq b_{D_f, f} - \sum_{w \in D_f} s_{w,f}
\end{align*}
hold, with a strict inequality holding for at least one member in $D \cup \{ f \}$. Here each worker $w$ is assigned to firm $f_w$, and a set of workers $D_f$ is assigned to each firm $f$ under the assignment $\Xb$.
\end{definition}
We reiterate here that the key difference between our definition of stability and that proposed by Kelso and Crawford~\cite{kelso1982job} is that in our setting firms can only hire a feasible set of workers $D \in \T_f \cup \emptyset$ rather than any set of workers, as in Kelso and Crawford~\cite{kelso1982job}. We note that we could also have considered other natural notions of stability, e.g., where firms must all match with a feasible set of workers and do not have the option of remaining unmatched unless $\emptyset \in \T_f$~\cite{jobMatchingConstraints2020}. In Section~\ref{sec:lowerBoundGeneral}, we consider such an alternative notion of stability, as in Kojima et al.~\cite{jobMatchingConstraints2020}, and present a generalization of our results under this stability notion in Appendix~\ref{apdx:r-stability-extensions}.

When studying stable arrangements $(\Xb, \{ \s_f \}_{f \in F})$, we are particularly interested in their \emph{total match value} $U$, i.e., the cumulative payoffs of all firms and workers combined, which is given by

\begin{equation} \label{eq:totalMatchRelation}
\begin{split}
    U(\Xb, \{ \s_f \}_{f \in F}) &= \sum_{w \in W} u_w(\Xb, \{ \s_f \}_{f \in F}) + \sum_{f \in F} v_f(\Xb, \{ \s_f \}_{f \in F}), \\
    &= \sum_{w \in W} (a_{w,f_w} + s_{w,f_w}) + \sum_{f \in F} (b_{D_f, f} - \sum_{w \in D_f} s_{w,f}), \\
    &= \sum_{w \in W} a_{w,f_w} + \sum_{f \in F} b_{D_f, f}, \\
    &= U(\Xb),
\end{split}
\end{equation}
where, with slight abuse of notation, we denote the total match value $U$ only as a function of the corresponding assignment. We do so, since the salaries do not influence the total match value of an arrangement. We now present the notion of efficiency that we use in this work.

\begin{definition} [Efficiency of Assignments] \label{def:efficiency}
We define an assignment $\Xb^*$ of workers to firms to be \emph{efficient} if it maximizes the total match value among the class of all feasible assignments, i.e., $U(\Xb^*) \geq U(\Xb)$ for all feasible assignments $\Xb$.
\end{definition}

\subsection{Efficiency of Stable Arrangements} \label{subsec:efficiency}

In this section, we present a consequence of stability on the efficiency of the resulting assignment. In particular, we establish that the assignment corresponding to any stable arrangement is efficient, i.e., it achieves the highest total match value among all feasible assignments.

\begin{theorem} [Efficiency of Stable Arrangements] \label{thm:stableOptimal}
If an arrangement $(\Xb^*, \{ \s_f \}_{f \in F})$ is stable, then the assignment $\Xb^*$ is efficient.
\end{theorem}

While stable arrangements may, in general, not exist (see Section~\ref{sec:nonExistence}), Theorem~\ref{thm:stableOptimal} establishes that if the market works at all, i.e., if stable arrangements exist, then the market works well as it results in efficient assignments. To prove Theorem~\ref{thm:stableOptimal}, we leverage the following necessary and sufficient condition for an arrangement to be stable.

\begin{lemma} [Necessary and Sufficient Condition for Stability] \label{lem:NSC-Stability}
Let $\u = (u_w)_{w \in W}$ and $\v = (v_f)_{f \in F}$ represent the payoff vectors of the workers and the firms under the arrangement $(\Xb^*, \{ \s_f \}_{f \in F})$. Then, an arrangement $(\Xb^*, \{ \s_f \}_{f \in F})$ is stable if and only if it is feasible, individually rational for workers and firms, and it holds for each feasible set of workers $D \in \T_f \cup \emptyset$ for each firm $f \in F$ that $\sum_{w \in D} a_{w,f} + b_{D,f} \leq \sum_{w \in D} u_w + v_f$, i.e., the sum of the payoffs of a firm $f$ and any feasible set of workers $D$ is at least the sum of their match values $\sum_{w \in D} a_{w,f} + b_{D,f}$.
\end{lemma}

For a proof of Lemma~\ref{lem:NSC-Stability}, see Appendix~\ref{apdx:pfnsc-stability}. Lemma~\ref{lem:NSC-Stability} characterizes a necessary and sufficient condition such that no feasible coalition of workers and a firm can improve upon their payoffs in a manner that would make each agent in the coalition weakly better off and at least one agent in the coalition strictly better off. We now leverage Lemma~\ref{lem:NSC-Stability} to prove Theorem~\ref{thm:stableOptimal}.

\begin{proof}
To prove this claim, we show that the assignment $\Xb^*$ 
has the highest total match value, i.e., $U(\Xb^*) \geq U(\Xb)$ for any other feasible assignment $\Xb$. To this end, let $\u = (u_w)_{w \in W}$ and $\v = (v_f)_{f \in F}$ be the payoffs of the workers and the firms, respectively, under the stable arrangement $(\Xb^*, \{ \s_f \}_{f \in F})$. Then, by Lemma~\ref{lem:NSC-Stability} it follows for each firm $f$ and feasible set of workers $D \in \T_f$ that 
\begin{align} \label{eq:stabilityConsequence}
    \sum_{w \in D} u_w + v_f \geq b_{D,f} + \sum_{w \in D} a_{w,f}.
\end{align}

Next, consider an efficient assignment $\Yb$, where $P_f^*$ denotes the set of workers matched to firm $f$ or the empty set in case no workers are matched to firm $f$ under the efficient assignment. Then, it follows that the total match value of the efficient assignment is given by
\begin{align*}
    U(\Yb) = \sum_{f \in F} \Bigg( \sum_{w \in P_f^*} a_{w,f} + b_{P_f^*,f} \Bigg) = \sum_{f \in F: P_f^* \neq \emptyset} \Bigg( \sum_{w \in P_f^*} a_{w,f} + b_{P_f^*,f} \Bigg).
\end{align*}
Next, for each set of workers $P_f^* \neq \emptyset$ it follows by Equation~\eqref{eq:stabilityConsequence} that
\begin{align*}
    \sum_{w \in P_f^*} a_{w,f} + b_{P_f^*,f} \leq \sum_{w \in P_f^*} u_w + v_f, \quad \text{for all } f \in F \text{ such that } P_f^* \neq \emptyset.
\end{align*}
Summing this inequality over all firms that are matched under the assignment $\Yb$ and observing by the feasibility of the assignment $\Yb$ that each worker $w$ belongs to at most one set $P_f^*$, i.e., each worker is matched to at most one firm under the efficient assignment, we obtain that
\begin{align*}
    U(\Yb) &= \sum_{f \in F: P_f^* \neq \emptyset} \Bigg( \sum_{w \in P_f^*} a_{w,f} + b_{P_f^*,f} \Bigg) \leq \sum_{f \in F: P_f^* \neq \emptyset} \Bigg( \sum_{w \in P_f^*} u_w + v_f \Bigg) \stackrel{(a)}{\leq} \sum_{w \in W} u_w + \sum_{f \in F} v_f \stackrel{(b)}{=} U(\Xb^*),
\end{align*}
where (a) follows from the individual rationality of workers and firms, and (b) follows from the sequence of equalities in Equation~\eqref{eq:totalMatchRelation}. This proves the efficiency of the assignment $\Xb^*$.
\end{proof}
Theorem~\ref{thm:stableOptimal} establishes that if a stable arrangement exists, then the corresponding assignment has a total match value that is the highest across all feasible assignments. We note that this result generalizes several existing results on the efficiency of competitive equilibria to the setting of two-sided matching markets with transfers and distributional constraints. In particular, Theorem~\ref{thm:stableOptimal} generalizes the efficiency of competitive equilibria in matching markets without constraints~\cite{bikhchandani1997competitive}, the efficiency of stable matchings in two-sided markets subject to distributional constraints but without transfers~\cite{kamada2015efficient}, and the efficiency of stable arrangements in one-to-one matching markets~\cite{shapley1971assignment}.

\section{Existence of Stable Arrangements under General Constraints and Preferences} \label{sec:oneFirmExistence}

The substitutes condition is often critical for the stability of the corresponding market outcome in both matching markets without constraints~\cite{kelso1982job,hatfield2005matching} and matching with constraints~\cite{jobMatchingConstraints2020} in the transferable utility setting. In contrast to earlier works that use the substitutes condition to establish the existence of stable arrangements, in this section, we study the question of whether stable arrangements can be guaranteed to exist under a general class of firm preferences and constraint structures beyond those that preserve the substitutes condition. To this end, we first present an 
example to demonstrate, even when the firm's preferences are linear over the workers, that the substitutes condition may not hold under general constraints (Section~\ref{subsec:vioSubsttutes}). Despite this result, we show in Section~\ref{subsec:oneFirmExistence} that stable arrangements are guaranteed to exist in a setting with one firm under a general class of firm preferences and constraint structures.

\subsection{Violation of Substitutes Condition} \label{subsec:vioSubsttutes}

In this section, we present an example where firms have linear preferences over workers to show that the substitutes condition may be violated under general constraints. To elucidate this example, we first define 
the substitutes condition, which states that an increase in the salary of a given worker weakly increases the demand (of a firm) for all other workers. To this end, we denote the salary vector of all workers other than worker $w$ as $\s_{-w,f}$, and the demand correspondence of a firm $f$ given a vector of salaries $\s_f$ and feasibility collection $\T_f$ by $\Yc_f(\s_f, \T_f) = \{D \in \T_f \cup \emptyset: D \in \argmax_{D' \in \T_f \cup \emptyset} \{ b_{D',f} - \sum_{w \in D'} s_{w,f} \} \}$. Then, the substitutes condition is as follows.

\begin{definition} [Substitutes Condition]
Suppose that the salary of a worker $w$ is increased from $s_{w,f}$ to $s_{w,f}^{\prime}$ for a given firm $f$. Then, the demand correspondence $\Yc_f$ of that firm satisfies the substitutes condition if for any set of workers $D \in \Yc_f$, there exists a set $D' \in \Yc_f((\s_{-w,f}, s_{w,f}^{\prime}), \T_f)$ such that $D \backslash \{w \} \subseteq D'$.
\end{definition}
We now show through a counterexample with one firm and three workers, where the firm has linear preferences over the workers, that raising the salary of one worker decreases the demand of that firm for another worker, thereby violating the substitutes condition. We refer to~\cite{jobMatchingConstraints2020} for further examples of constraint structures where the substitutes condition may not hold.

\begin{example} [Violation of Substitutes Condition] \label{eg:violationSubstitutes}

Consider a setting of one firm $f_1$ and three workers $w_1, w_2, w_3$, where the firm has the following upper bound hiring constraints: (i) at most one of workers $w_1$ and $w_2$ can be hired, i.e., $x_{w_1, f_1} + x_{w_2, f_1} \leq 1$, and (ii) at most one of workers $w_2$ and $w_3$ can be hired, i.e., $x_{w_2, f_1} + x_{w_3, f_1} \leq 1$, as depicted in Figure~\ref{fig:substitutes} in Appendix~\ref{apdx:substitutes-violation}. Note that under these constraints, the feasibility collection $\T_{f_1} = \{ \emptyset, \{w_1 \}, \{ w_2 \}, \{w_3\}, \{w_1, w_3 \} \}$. Furthermore, we let the preferences of the firm be linear, where, recall that, $c_{w,f_1}$ represents the value of firm $f_1$ for worker $w$, and $a_{w,f_1}$ is the value of each worker $w$ for firm $f_1$, as given in Table~\ref{tab:1}.

\begin{table}[tbh!] 
\centering
\caption{\small \sf Value of workers for each firm and the firm for each worker in a one-firm, three-worker example where the substitutes condition does not hold.}
\vspace{10pt}
\footnotesize
\begin{tabular}{|c|c|}
\hline
\textbf{Value of Worker for Firm} & \textbf{Value of Firm for Worker} \\ \hline  $a_{w_1,f_1} = -0.5$                     & $c_{w_1,f_1} = 1.5$                         \\ \hline $a_{w_2,f_1} = -0.5$                     & $c_{w_2,f_1} = 2.5$                       \\ \hline $a_{w_3,f_1} = -0.5$                     & $c_{w_3,f_1} = 1.5$       \\ \hline
\end{tabular} 
\end{table} \label{tab:1}
We now show that the substitutes condition is violated by studying the demand correspondence of the firm under two different salary profiles. In particular, first consider the salary profile given by $s_{w_1,f_1} = 0.5$, $s_{w_2,f_1} = 1$ and $s_{w_3,f_1} = 0.5$. Under these salaries, the firm maximizes its payoff by hiring workers $w_1$ and $w_3$ resulting in a total payoff of two. Thus, the optimal demand correspondence of the firm at this salary profile $\s_{f_1} = (0.5, 1, 0.5)$ is given by $\Yc_{f_1}(\s_{f_1}, \T_{f_1}) = \{ \{ w_1, w_3 \} \}$.

Next, if the salary of worker one is increased to $s_{w_1,f_1}' = 1.1$, then the firm maximizes its payoff by hiring worker $w_2$, resulting in a payoff of 1.5. Thus, the optimal demand correspondence under this new salary profile $\s_{f_1}' = (1.1, 1, 0.5)$ is $\Yc_{f_1}(\s_{f_1}', \T_{f_1}) = \{ \{ w_2 \} \}$. However, worker $w_3$ is not in the optimal demand correspondence of the firm under this new salary profile, i.e., $w_3 \notin \{ w_2 \}$. Thus, the demand correspondence of the firm does not satisfy the substitutes condition. \qed
\end{example}

This example shows that for two-sided matching markets subject to distributional constraints the substitutes condition may be violated even under linear preferences for relatively small problem instances. As a result, even in such simplified settings, arguments relying on the satisfaction of the substitutes condition of the firm's demand correspondence cannot be used to guarantee the existence of stable arrangements.

\subsection{Existence of Stable Arrangements for a One-Firm Setting} \label{subsec:oneFirmExistence}

While Example~\ref{eg:violationSubstitutes} indicates that the substitutes condition may not hold, even under linear preferences, for general constraint structures, in this section, we show in a one-firm setting that stable arrangements exist irrespective of the nature of the constraint structure or the preferences of the firm $f$ over workers.

\begin{theorem} [Existence of Stable Arrangement for One Firm] \label{thm:ExistenceOneFirm}
In a one-firm setting, there exists a stable arrangement under any feasibility collection $\T_f$ and any vector of valuations $\b = (b_{D,f})_{D \in \T_f}$ of the firm $f$.
\end{theorem}

\begin{proof}
To prove this claim, we construct an arrangement $(\Xb^*, \s_f^*)$ and show that this arrangement is stable. For the remainder of this proof, we drop the subscript $f$ for the ease of notation since we are considering the setting of just one firm.

To define the arrangement $(\Xb^*, \s^*)$, consider the following problem
\begin{align*}
    D^* = \argmax_{D \in \T \cup \emptyset} \Bigg\{ \max_{\genfrac{}{}{0pt}{2}{\s: s_{w} + a_{w} \geq 0,}{\forall w \in D} }  b_{D} - \sum_{w \in D} s_{w} \Bigg\}.
\end{align*}
First observe that the solution of the inner maximization problem must satisfy $s_w = -a_w$, and thus, it follows that the above problem can be rewritten as
\begin{align} \label{eq:combinatorial}
    D^* = \argmax_{D \in \T \cup \emptyset} \Bigg\{ b_D + \sum_{w \in D} a_w \Bigg\}.
\end{align}
Now, consider the assignment matrix $\Xb^*$ that encodes the assignment of workers in the set $D^*$ to the firm, and let the salaries  $\s^*$ be such that $s_w^* = -a_w$ for $w \in D^*$ and any $s_w^* < -a_w$ for $w \notin D^*$. Then, we claim that the arrangement $(\Xb^*, \s^*)$ is stable.

To see this, first observe that such an arrangement is individually rational for all workers since the payoff for matched workers $w \in D^*$ is given by $a_w + s_w^* \geq 0$, while all unmatched workers obtain a payoff of zero. Next, the arrangement $(\Xb^*, \s^*)$ is individually rational for the firm, since the payoff of the firm under the arrangement is at least its payoff if unmatched since $D^* \subseteq \T \cup \emptyset$. Furthermore, note that the assignment $\Xb^*$ is feasible since the set of workers $D^* \subseteq \T \cup \emptyset$. Finally, to prove stability we claim that there is no set of workers $D \in \T \cup \emptyset$ with corresponding salaries $\Tilde{\s}$ such that (i) all workers in the set $D$ are weakly better off than under the arrangement $(\Xb^*, \s^*)$, and (ii) the firm is at least as well of as under the assignment $D^*$ with salaries $\s^*$, i.e., $b_D - \sum_{w \in D} s_w^* \geq b_{D^*} - \sum_{w \in D^*} \Tilde{s}_w$, with at least one of the inequalities in (i) or (ii) being strict.

To establish this claim, we proceed by contradiction. In particular, suppose that there is a vector of salaries $\Tilde{\s}$ and a feasible set of workers $\Tilde{D} \in \T \cup \emptyset$ such that there is a worker $w' \in \Tilde{D}$ that is strictly better off than under the arrangement $(\Xb^*, \s^*)$. Furthermore, suppose that the firm and all other workers $w \in \Tilde{D} \backslash \{ w' \}$ are weakly better off than under the arrangement $(\Xb^*, \s^*)$. That is, worker $w'$ receives a strictly positive payoff, i.e., $\Tilde{s}_{w'} > -a_{w'}$, while all other workers in $\Tilde{D}$ receive a non-negative payoff, i.e., $\Tilde{s}_{w} \geq -a_{w}$ for all workers $w \in \Tilde{D} \backslash \{ w' \}$. Note that the case where the firm is strictly better off is not possible since for any set of workers $\Tilde{D} \in \T \cup \emptyset$ the payoff of the firm under the matching with workers in $D^*$ is at least that under any other feasible assignment by the definition of $D^*$. That is, the firm's payoff $v$ under the arrangement $(\Xb^*, \s^*)$ satisfies
\begin{align*} 
    v(\Xb^*, \{\s^*\}) = b_{D^*} - \sum_{w \in D^*} s_w^* \geq b_{\Tilde{D}} - \sum_{w \in \Tilde{D}} \Tilde{s}_w
\end{align*}
for any vector of incomes $\Tilde{\s}$ such that $\Tilde{s}_w+a_w \geq 0$ and all sets of workers $\Tilde{D}$ by the definition of $D^*$.

Now, since worker $w'$ receives a strictly positive payoff while all other workers in $\Tilde{D}$ receive a non-negative payoff it follows that
\begin{align} \label{eq:helper2}
    b_{\Tilde{D}} - \sum_{w \in \Tilde{D}} \Tilde{s}_w < b_{\Tilde{D}} + \sum_{w \in \Tilde{D}} a_w.
\end{align}
Finally, denoting $\Tilde{\Xb}$ as the assignment of workers in the set $\Tilde{D}$ to the firm and $\Tilde{v}$ as the payoff of the firm under the arrangement $(\Tilde{\Xb}, \Tilde{\s})$, we get that
\begin{align*}
    \Tilde{v}(\Tilde{\Xb}, \Tilde{\s}) = b_{\Tilde{D}} - \sum_{w \in \Tilde{D}} \Tilde{s}_w \stackrel{(a)}{<} b_{\Tilde{D}} + \sum_{w \in \Tilde{D}} a_w \stackrel{(b)}{\leq} b_{D^*} + \sum_{w \in D^*} a_w \stackrel{(c)}{=} b_{D^*} - \sum_{w \in D^*} s_w^* = v(\Xb^*, \s^*),
\end{align*}
where (a) follows by Equation~\eqref{eq:helper2}, (b) follows from the definition of $D^*$, and (c) follows from the definition of $\s^*$. This relation implies that if there is an arrangement that leads to weakly higher payoffs for all workers with one worker receiving a strictly positive payoff, then the firm will necessarily obtain a strictly lower payoff than under the arrangement $(\Xb^*, \s^*)$. Thus, we have arrived at our desired contradiction, which proves that the arrangement $(\Xb^*, \s^*)$ is stable.
\end{proof}
Theorem~\ref{thm:ExistenceOneFirm} is particularly striking since it establishes that stable arrangements exist under arbitrary constraint structures and agent preferences, including those under which the substitutes condition may not hold, as elucidated in Example~\ref{eg:violationSubstitutes}. In particular, this result guarantees the existence of stable arrangements when the necessary and sufficient conditions for the substitutes condition in Kojima et al.~\cite{jobMatchingConstraints2020} do not hold. We mention that this result also extends naturally under another notion of stability (see Section~\ref{sec:lowerBoundGeneral}), wherein we relax the individual rationality of firms, as in Kojima et al.~\cite{jobMatchingConstraints2020}, and we present this extension in Appendix~\ref{apdx:r-stability-extensions}. However, this result on the existence of stable arrangements under arbitrary preferences and constraints does not extend to the setting of multiple firms (see Section~\ref{sec:nonExistence}).

\section{Stable Arrangements under Linear Preferences and Hiring Quotas} \label{sec:multiFirms}


In the previous section, we established that stable arrangements exist under general constraints and agent preferences in the setting of a single firm. In this section, we consider the multiple firm setting and focus on matching markets where firms have linear preferences over workers and upper and lower bound distributional constraints. Our motivation for studying this class of markets is two-fold. First, linear preferences have been studied extensively in matching markets~\cite{bikhchandani1997competitive} and serve as a direct analog to the matching market with additively separable preferences considered by Crawford and Knoer~\cite{crawford1981job}. Next, upper and lower bound constraints on hiring are natural in many applications~\cite{bckm} and have been widely studied in two-sided matching markets without transfers~\cite{kamada2015efficient,kamada2017stability,kamada2018stability} and in two-sided matching markets with transfers~\cite{jobMatchingConstraints2020}. Notably, Kojima et al.~\cite{jobMatchingConstraints2020} show that a sub-class of upper and lower bound hiring constraints are necessary and sufficient for the preservation of the substitutes condition, which guarantees the existence of stable arrangements.

We begin our study of this class of markets by establishing in Section~\ref{sec:nonExistence} that stable arrangements are not guaranteed to exist even in a setting with two firms. This result on the non-existence of stable arrangements makes the study of this class of markets with linear preferences particularly relevant. As a result, we then study the conditions on the constraint structure under which stable arrangements exist and can be computed efficiently. To this end, we introduce a class of constraint structures in Section~\ref{sec:constraintProp} under which the substitutes condition holds and use linear programming duality to establish the existence of and efficiently compute stable arrangements in Sections~\ref{sec:MultipleFirmExistence}-\ref{sec:mainExistence}. We mention here that our main contribution in this section is to provide a method to efficiently compute stable arrangements through a connection between linear programming duality and the existence of stable arrangements. For the ease of exposition, we focus on the setting with upper bound hiring constraints and present a generalization of our results under lower bound constraints in Appendix~\ref{sec:lowerBoundGeneral}. In the context of both upper and lower bound hiring constraints, we note that the condition on the constraint structure that we require to establish the existence of stable arrangements using linear programming is analogous to Kojima et al.'s~\cite{jobMatchingConstraints2020} necessary and sufficient condition for the preservation of the substitutes condition. We refer to Remark~\ref{rmk:gpc} in Appendix~\ref{apdx:pfLB} for a distinction between the sets of constraints we consider and those in Kojima et al.~\cite{jobMatchingConstraints2020}, and for a description on how our approach extends to the class of constraints considered in Kojima et al.~\cite{jobMatchingConstraints2020}.

\subsection{Stable Arrangements Need not Exist} \label{sec:nonExistence}

While stable arrangements are guaranteed to exist in the setting with one firm (Theorem~\ref{thm:ExistenceOneFirm}), we now present an example of a market with multiple firms where a stable arrangement does not exist. In particular, we provide an example of a market where the firms have upper bound constraints on hiring and linear preferences over workers.

\begin{proposition} [Non-Existence of Stable Arrangements] \label{prop:nonExistenceStable}
In the setting with upper bound constraints on hiring and where firms have linear preferences over workers, there exists a market instance with three workers and two firms such that no stable arrangement exists.
\end{proposition}

\begin{proof}
We first formally define a market instance with two firms, labelled $f_1$ and $f_2$, and three workers, labelled $w_1$, $w_2$, and $w_3$, and their corresponding preferences and constraints on hiring. In particular, suppose that the firm $f_1$ has the following upper bound constraints on hiring: (i) at most one of workers $w_1$ and $w_2$ can be hired, i.e., $x_{w_1, f_1} + x_{w_2, f_1} \leq 1$, and (ii) at most one of workers $w_2$ and $w_3$ can be hired, i.e., $x_{w_2, f_1} + x_{w_3, f_1} \leq 1$. Furthermore, suppose that the firm $f_2$ has the following upper bound constraints on hiring: (i) at most one of workers $w_2$ and $w_3$ can be hired, i.e., $x_{w_2, f_2} + x_{w_3, f_2} \leq 1$, and (ii) at most one of workers $w_1$ and $w_3$ can be hired, i.e., $x_{w_1, f_2} + x_{w_3, f_2} \leq 1$. These constraint structures are depicted in Figure~\ref{fig:non-ex-stability} in Appendix~\ref{apdx:stablenonEx}.

Next, consider any set of values of each firm $f$ for each worker $w$, given by $c_{w,f}$, and values of each worker $w$ for each firm $f$, given by $a_{w,f}$, that satisfy the match values in Table~\ref{tab:matchValNonExistence}. Recall that the match value of the worker-firm pair $(w,f)$ is given by $\alpha_{w,f} = a_{w,f} + c_{w,f}$.

\begin{table}[tbh!]
\centering
\caption{\small \sf Match Values for two-firm, three-worker instance to show that stable arrangements are not guaranteed to exist.}
\vspace{10pt}
\footnotesize
\begin{tabular}{|c|c|}
\hline
\textbf{Match Value for Firm $f_1$} & \textbf{Match Value for Firm $f_2$} \\ \hline  $\alpha_{w_1,f_1} = 0.9$                     & $\alpha_{w_1,f_2} = 0.8$                          \\ \hline $\alpha_{w_2,f_1} = 1.1$                 & $\alpha_{w_2,f_2} = 1$                         \\ \hline $\alpha_{w_3,f_1} = 1$  & $\alpha_{w_3,f_2} = 1.1$                        \\ \hline
\end{tabular}
\label{tab:matchValNonExistence}
\end{table}

We now show for the above defined market instance that stable arrangements are not guaranteed to exist. To this end, we suppose for contradiction that a stable arrangement exists and establish that the payoff vectors $\v$ of the firms and $\u$ of workers satisfy the following two incompatible relations: (i) $u_{w_2} + u_{w_3} + v_{f_1} + v_{f_2} \leq 2.1$, and (ii) $u_{w_2} + u_{w_3} + v_{f_1} + v_{f_2} \geq 2.2$.

We begin by showing relation (i). To this end, note that if a stable arrangement $(\Xb^*, \{\s_f \}_{f \in F})$ exists, then, by Theorem~\ref{thm:stableOptimal}, the assignment $\Xb^*$ corresponding to such an arrangement must be optimal. We note in this market setting that there is a unique optimal assignment, wherein workers $w_1$ and $w_3$ are assigned to firm $f_1$ and worker $w_2$ is assigned to firm $f_2$ resulting in a total match value of $2.9$. Under this optimal assignment and corresponding salaries $\{\s_f \}_{f \in F}$, the sum of the payoffs of firm $f_1$ and the workers $w_1, w_3$ must satisfy the following relation:
\begin{equation} \label{eq:f1Payoff}
\begin{split}
    u_{w_1} + u_{w_3} + v_{f_1} &= a_{w_1,f_1} + s_{w_1,f_1} + a_{w_3,f_1} + s_{w_3, f_1} + c_{w_1,f_1} + c_{w_3,f_1} - s_{w_1,f_1} - s_{w_3, f_1}, \\
    &= \alpha_{w_1,f_1} + \alpha_{w_3,f_1}.
\end{split}
\end{equation}
In addition, the following relation on the payoff of firm $f_2$ and the worker $w_2$ must hold under the arrangement $(\Xb^*, \{\s_f \}_{f \in F})$:
\begin{align} \label{eq:f2Payoff}
    &u_{w_2} + v_{f_2} = \alpha_{w_2,f_2}.
\end{align}

Next, since the arrangement $(\Xb^*, \{\s_f \}_{f \in F})$ is stable, it must be that $u_{w_1} \geq \alpha_{w_1,f_2}$. To see this, note that if $u_{w_1} < \alpha_{w_1,f_2}$ then worker $w_1$ would rather be matched with firm $f_2$, with firm $f_2$ receiving a non-negative payoff under this matching and thus resulting in a profitable deviation, thereby violating the stability of $(\Xb^*, \{\s_f \}_{f \in F})$.

Then, combining Equation~\eqref{eq:f1Payoff} and the observation that $u_{w_1} \geq \alpha_{w_1,f_2}$, it follows that
\begin{align*}
    u_{w_3} + v_{f_1} \leq \alpha_{w_1,f_1} + \alpha_{w_3,f_1} - \alpha_{w_1,f_2}.
\end{align*}
Next, summing the above inequality with Equation~\eqref{eq:f2Payoff}, we obtain that
\begin{align} \label{eq:condition1}
    u_{w_2} + u_{w_3} + v_{f_1} + v_{f_2} \leq \alpha_{w_1,f_1} + \alpha_{w_3,f_1} - \alpha_{w_1,f_2} + \alpha_{w_2,f_2} = 2.1,
\end{align}
which establishes relation (i).

We now establish the second relation that $u_{w_2} + u_{w_3} + v_{f_1} + v_{f_2} \geq 2.2$ under the arrangement $(\Xb^*, \{\s_f \}_{f \in F})$. To this end, observe that worker $w_2$ is feasible for firm $f_1$ and worker $w_3$ is feasible for firm $f_2$. Then, by the necessary and sufficient condition for stability, established in Lemma~\ref{lem:NSC-Stability}, it must hold that $u_{w_2} + v_{f_1} \geq \alpha_{w_2,f_1}$, and that $u_{w_3} + v_{f_2} \geq \alpha_{w_3,f_2}$. Summing these inequalities gives
\begin{align} \label{eq:condition2}
    u_{w_2} + u_{w_3} + v_{f_1} + v_{f_2} \geq \alpha_{w_2,f_1} + \alpha_{w_3,f_2} = 2.2.
\end{align}
However, we note that Equations~\eqref{eq:condition1} and~\eqref{eq:condition2} cannot both be simultaneously satisfied since $\alpha_{w_2,f_1} + \alpha_{w_3,f_2} = 2.2 > 2.1 = \alpha_{w_1,f_1} + \alpha_{w_3,f_1} - \alpha_{w_1,f_2} + \alpha_{w_2,f_2}$. Thus, we have our desired contradiction, which proves that a stable arrangement cannot exist in the above described market instance.
\end{proof}
Proposition~\ref{prop:nonExistenceStable} shows that even when firms have linear preferences over workers, there are constraint structures, which include only upper bound constraints on hiring, that may preclude the existence of stable arrangements. This result provides an example of a market instance where stable arrangements do not exist in the transferable utility setting, which is in contrast to the work of Kamada and Kojima~\cite{kamada2017stability}, who provide an example of constraints under which stable matchings do not exist for the non-transferable utility setting. We also note that Proposition~\ref{prop:nonExistenceStable} differs from prior work~\cite{jobMatchingConstraints2020} on matching markets with transfers under distributional constraints. In particular, we provide an example, through Proposition~\ref{prop:nonExistenceStable}, when stable arrangements do not exist, while Kojima et al.~\cite{jobMatchingConstraints2020} provide examples of constraint structures under which the substitutes condition does not hold, which does not necessarily prevent the existence of stable arrangements (see Theorem~\ref{thm:ExistenceOneFirm}). Thus, Proposition~\ref{prop:nonExistenceStable} motivates a deeper study of the class of constraints under which stable arrangements exist when firms have linear preferences over workers.

\subsection{Constraint Structures that Guarantee Existence of Stable Arrangements} \label{sec:constraintProp}

In the previous section, we presented an example where stable arrangements did not exist under general upper bound constraints on hiring. As a result, in this section, we present classes of constraints under which the substitutes condition holds, which guarantees the existence of stable arrangements. In particular, we introduce \emph{polymatroids} and \emph{hierarchical} constraint structures, under which Kojima et al.~\cite{jobMatchingConstraints2020} show that the substitutes condition is preserved. We reiterate here that we begin with the study of only upper bound constraints on hiring and present the generalization of our results to the setting with lower bound constraints in Section~\ref{sec:lowerBoundGeneral}.

We now introduce the notions of a polymatroid and a hierarchy. To this end, we begin by recalling that in the setting with upper bound constraints on hiring, an assignment $\Xb$ is \emph{feasible} if each worker is assigned to at most one firm and for all firms $f$ it holds that $\sum_{w \in D} x_{w,f} \leq \Bar{\lambda}_D$ for all sets of workers $D \in \H_f$. While under general upper bound hiring constraints, stable arrangements do not exist (Proposition~\ref{prop:nonExistenceStable}), polymatroids and hierarchical constraint structures impose certain restrictions on the constraint set $\H_f$ and quotas $\llambda_f$ for each firm $f$ such that the substitutes condition holds. 

To elucidate the types of restrictions on the constraint sets and quotas in the case of polymatroids, we first introduce the definitions of \emph{intersecting subsets}, an \emph{intersecting family} and \emph{submodular set functions}. In particular, two subsets $A,B$ of a finite set $S$ are defined to be \emph{intersecting} if none of the sets $A \cap B$, $A \backslash B$ and $B \backslash A$ are empty. A family $\H'$ of subsets is \emph{intersecting} over the set $S$ if $\emptyset \notin \H'$ and for all intersecting subsets $A,B \in S$ it holds that $A \cap B, A \cup B \in \H'$. Finally, we define a set function $g$ to be \emph{submodular}\footnote{We mention that a function $g$ is \emph{supermodular} if $-g$ is submodular.} on $A, B$ if it holds that $g(A) + g(B) \geq g(A \cup B) + g(A \cap B)$. Then, we can define a polymatroid, as in Frank~\cite{frank1984generalized}, as follows.

\begin{definition} [Polymatroid~\cite{frank1984generalized}] \label{def:polyMatroid}
Let $\H'$ be an intersecting family over a set $S$ and $g$ be an integer-valued function on $\H'$ that is submodular on intersecting sets. Then, the polyhedron
\begin{align*}
    P = \{ \y: \y \geq \0, \sum_{d \in D} y_d \leq g(D), \text{ for } D \in \H' \}
\end{align*}
defines a polymatroid $P(\H', g)$, and the corresponding constraints are polymatroid constraints.
\end{definition}
Note that in the context of matching markets, the set $S = W \times F$, the family $\H' =  \cup_{f \in F} \left( \H_f \times \{ f \} \right)$ corresponds to the set of upper bound hiring constraints that satisfy the property of an intersecting family, and the submodular function $g$ imposes a restriction on the upper bound hiring quotas $\Bar{\llambda}_f$. 

An important class of an intersecting family (of constraints) $\H'$ is a hierarchy, which is a family of subsets such that any two members of $\H'$ are either disjoint or one is a subset of the other.

\begin{definition} [Hierarchy] \label{def:hierarchy}
A family (of constraints) $\H'$ is a hierarchy if for every pair of elements $S, S' \in \H'$, either $S \subseteq S'$ or $S' \subseteq S$ or $S \cap S' = \emptyset$.
\end{definition}
Hierarchical constraints have been widely studied in two-sided matching markets~\cite{kamada2018stability,bilemi-2004,gul2019market} and are closely related to polymatroids, as is elucidated by the following remark.

\begin{remark} [Relation between Hierarchies and Polymatroids] \label{rmk:polyhier}
Hierarchical constraints are a special case of polymatroid constraints, i.e., for any non-negative integer-valued function $g'$ on a hierarchy $\H'$, the polyhedron $P = \{ \y: \y \geq \0, \sum_{d \in D} y_d \leq g'(D), \text{ for } D \in \H' \}$ is a polymatroid $P(\H', g')$. To see this, note that no two members of $\H'$ are intersecting subsets by the definition of a hierarchy, and thus (i) $\H'$ is an intersecting family, and (ii) the function $g'$ is submodular on intersecting sets.
\end{remark}
Having noted this fundamental connection between polymatroids and hierarchies, we present an example in the context of the two-sided matching markets studied in this work to elucidate these notions and refer to~\cite{bckm} for further examples of such constraints.

\begin{example} [Polymatroids and Hierarchical Constraint Structures] \label{eg:heirarchy}
We refer to Figure~\ref{fig:polyheir} in Appendix~\ref{eg:polyheirEg} for a depiction of the constraints considered in this example. In particular, consider a setting of one firm $f_1$ and four workers $w_1, w_2, w_3, w_4$. In this case, first consider the constraint structure where firm can hire (i) a total of at most two workers, (ii) at most one of workers $w_1$ and $w_2$, and (iii) at most one of workers $w_3$ and $w_4$.  Observe that this constraint structure is a hierarchy since the hiring constraints on the subsets of workers are given by (i) $D_1 = \{w_1, w_2, w_3, w_4 \}$, (ii) $D_2 = \{ w_1, w_2 \}$, and (iii) $D_3 = \{ w_3, w_4 \}$, which are either disjoint sets, i.e., $D_2 \cap D_3 = \emptyset$, or subsets of the other, i.e., $D_2, D_3 \subseteq D_1$. Also, by Remark~\ref{rmk:polyhier}, observe that the polyhedron corresponding to this constraint structure specifies a polymatroid. On the other hand, if constraint (iii) is replaced by the constraint that the firm can hire at most one of workers $w_2$ and $w_3$, then the constraint structure is no longer a hierarchy since the sets $D_2 = \{ w_1, w_2 \}$ and $D_{3}' = \{ w_2, w_3 \}$ are neither disjoint nor subsets of each other. The polyhedron corresponding to this constraint structure also does not specify a polymatroid since the associated family (of constraints) is not intersecting.
\qed
\end{example}

\subsection{Linear Programming and Stable Arrangements} \label{sec:MultipleFirmExistence}

In this section, we present our main result on the connection between linear programming and the existence of stable arrangements. In particular, we use linear programming duality to establish that stable arrangements exist and can be computed efficiently under both polymatroid and hierarchical constraints.

We now present the main result of this section (Theorem~\ref{thm:ExistenceHierarchy}), which uses linear programming duality to establish the existence of stable arrangements under polymatroid constraints.

\begin{theorem} [Linear Programming and Stable Arrangements] \label{thm:ExistenceHierarchy}
Suppose that for each firm $f$ the constraint structure $\H_f \times \{ f\}$ is an intersecting family over $W \times F$ and there is some integer-valued function $g_f$ on $\H_f \times \{f\}$ that is submodular on intersecting sets. Further, suppose that the upper bound hiring quotas $\Bar{\llambda}_f$ satisfy $\Bar{\lambda}_{D} = g_f(D \times \{f\})$ for all $D \in \H_f$ and firms $f \in F$. Then, there exists a stable arrangement and this stable arrangement can be computed, using linear programming, in polynomial time in the number of firms, workers, and the cardinality of the constraint set given by $\sum_{f \in F} |\H_f|$.
\end{theorem}
Note here that the condition on the constraint structure and quotas in the statement of Theorem~\ref{thm:ExistenceHierarchy} correspond to polymatroid constraints (see Definition~\ref{def:polyMatroid}). Before proving this theorem, we first present its immediate corollary that there exists a stable arrangement, and this arrangement can be efficiently computed if the family (of constraints) $\H = \cup_{f \in F} (\H_f  \times \{f\} )$ forms a hierarchy.

\begin{corollary} [Linear Programming and Stable Arrangements] \label{cor:ExistenceHierarchy}
Suppose that the family (of constraints) $\H$ is a hierarchy. Then, for any non-negative integer-valued upper bound hiring quotas $\Bar{\llambda}_f$ for each firm $f \in F$, there exists a stable arrangement. Furthermore, this stable arrangement can be computed, using linear programming, in polynomial time in the number of firms, workers, and the cardinality of the constraint set given by $\sum_{f \in F} |\H_f|$.
\end{corollary}
Corollary~\ref{cor:ExistenceHierarchy} is an immediate consequence of Theorem~\ref{thm:ExistenceHierarchy} since hierarchical constraints are a special case of polymatroid constraints (see Remark~\ref{rmk:polyhier}). We now provide an outline of the proof of Theorem~\ref{thm:ExistenceHierarchy}, and present the corresponding details in Sections~\ref{sec:LP} and~\ref{sec:mainExistence}.

\begin{hproof}
To prove Theorem~\ref{thm:ExistenceHierarchy}, we need to show that there exists some stable arrangement $(\Xb^*, \{s_f \}_{f \in F})$ that can be computed via a linear program. Furthermore, the number of decision variables and constraints of this linear program must be polynomial in the number of firms, workers, and the cardinality of the constraint set given by $\sum_{f \in F} |\H_f|$. To compute the assignment $\Xb^*$, we note by Theorem~\ref{thm:stableOptimal} that it must be efficient, i.e., it must maximize the total match value. Thus, we formulate a linear program to compute efficient fractional assignments and show that this linear program has an integral optimal solution if the constraint structure corresponds to a polymatroid (Section~\ref{sec:LP}). We mention that the number of decision variables and constraints of this linear program is polynomial in the number of firms, workers, and the cardinality of the constraint set. Then, in Section~\ref{sec:mainExistence}, we use linear programming duality to establish that there exist salaries $\{\s_f \}_{f \in F}$ such that the payoffs of agents under the arrangement $(\Xb^*, \{\s_f \}_{f \in F})$ satisfy the necessary and sufficient condition for stability established in Lemma~\ref{lem:NSC-Stability}, which proves our claim.
\end{hproof}

\paragraph{Significance of Theorem~\ref{thm:ExistenceHierarchy}}
Theorem~\ref{thm:ExistenceHierarchy} establishes that under certain conditions on the constraint structure, which hold in many practical applications~\cite{bckm}, stable arrangements exist and can be computed efficiently using linear programming duality in the setting with multiple firms. In particular, our linear programming approach generalizes the duality theory approach of Shapley and Shubik~\cite{shapley1971assignment} in the one-to-one matching setting to the many-to-one matching setting under upper bound hiring constraints. Furthermore, together with Theorem~\ref{thm:stableOptimal}, Theorem~\ref{thm:ExistenceHierarchy} implies that if the upper bound hiring constraints are polymatroid constraints, then a stable arrangement can be realized that attains the highest total match value among all feasible assignments.

We reiterate here that the main contribution of Theorem~\ref{thm:ExistenceHierarchy} is to provide a method to compute stable arrangements efficiently through a connection between linear programming and stability. As a result, the linear programming approach of Theorem~\ref{thm:ExistenceHierarchy} provides an alternative perspective to and builds on the work of Kojima et al.~\cite{jobMatchingConstraints2020}, who established that polymatroid constraints guarantee the existence of stable arrangements, albeit under a different notion of stability wherein firms do not have the option of remaining unmatched. In particular, Theorem~\ref{thm:ExistenceHierarchy} not only provides an alternate proof to that in Kojima et al.~\cite{jobMatchingConstraints2020} for the existence of stable arrangements under polymatroid constraints but also shows that such arrangements can be computed efficiently using linear programming.

Finally, we mention that we consider the setting of lower bound hiring constraints in Section~\ref{sec:lowerBoundGeneral}, wherein we also establish the connection between linear programming duality and the existence of stable arrangements under the notion of stability in Kojima et al.~\cite{jobMatchingConstraints2020}.

\subsection{Linear Programming to Compute Efficient Assignments} \label{sec:LP}

In this section, we formulate a linear program to compute efficient fractional assignments and present a sufficient condition for this linear program to have an integral optimal solution. In particular, we show that if the upper bound hiring constraints are polymatroid constraints, then the optimal solution of this linear program corresponds to an efficient (integral) assignment.

We begin by presenting the following linear program, which is a fractional relaxation of the problem of computing efficient integral assignments under upper bound hiring constraints
\begin{maxi!}|s|[2]<b>
	{\Xb \in \mathbb{R}^{|W| \times |F|}}{U(\Xb) =\sum_{w \in W} \sum_{f \in F} \alpha_{w,f} x_{w,f}, \label{eq:ShapleyShubikObjFull}}
	{}
	{}
	\addConstraint{\sum_{f \in F} x_{w,f}}{\leq 1, \quad \forall w \in W \label{eq:SSWorkerAllocation},}
	\addConstraint{x_{w,f}}{ \geq 0, \quad \forall w \in W, f \in F \label{eq:SSnonnegativity-constraintsFull},}
	\addConstraint{ \sum_{w \in D} x_{w,f}}{ \leq \Bar{\lambda}_{D}, \quad \forall D \in \H_f, f \in F \label{eq:SSdistributionalUB-constraintsFull}.}
\end{maxi!}
Here~\eqref{eq:ShapleyShubikObjFull} represents the objective of maximizing the total match value of a fractional assignment,~\eqref{eq:SSWorkerAllocation} are worker assignment constraints,~\eqref{eq:SSnonnegativity-constraintsFull} are non-negativity constraints,  and~\eqref{eq:SSdistributionalUB-constraintsFull} represent the upper bound distributional constraints on hiring. We mention that we relaxed the binary allocation constraints with non-negativity Constraints~\eqref{eq:SSnonnegativity-constraintsFull}, which allows for fractional allocations. Other than the aforementioned fractional relaxation, the above linear program captures all the constraints required to ensure the feasibility of an assignment.

Even though the optimal solution of the linear Program~\eqref{eq:ShapleyShubikObjFull}-\eqref{eq:SSdistributionalUB-constraintsFull} may not be integral for general constraints (see Appendix~\ref{apdx:egNonIntegrality} for an example), we show that if the upper bound hiring constraints are polymatroid constraints, then its optimal solution is integral.

\begin{lemma} [Integral Optimal Solutions] \label{lem:integral-optimal-solutions}
Suppose that for each firm $f$ the constraint structure $\H_f \times \{ f\}$ is an intersecting family over $W \times F$ and there is some integer-valued function $g_f$ on $\H_f \times \{f\}$ that is submodular on intersecting sets. Further, suppose that the upper bound hiring quotas $\Bar{\llambda}_f$ satisfy $\Bar{\lambda}_{D} = g_f(D \times \{f\})$ for all $D \in \H_f$. Then, there is an optimal solution $\Xb^*$ of the linear Program~\eqref{eq:ShapleyShubikObjFull}-\eqref{eq:SSdistributionalUB-constraintsFull} such that $x_{w,f}^* \in \{ 0, 1 \}$ for all worker-firm pairs $(w,f) \in W \times F$.
\end{lemma}
The proof of Lemma~\ref{lem:integral-optimal-solutions} relies on the polymatroid intersection theorem~\cite{edmonds2003submodular}, which states that the optimal solution of a linear program under the intersection of two polymatroids is integral.

\begin{theorem} [Polymatroid Intersection Theorem~\cite{edmonds2003submodular}] \label{thm:polyintthm}
For a set $E$, a vector $\q$, and two polymatroids $P_1$ and $P_2$, the linear program $\max_{\y \in \mathbb{R}^{|E|}} \q^T \y \text{ s.t. } \y \in P_1 \cap P_2$ has an integral optimal solution.
\end{theorem}
Using Theorem~\ref{thm:polyintthm}, we complete the proof of Lemma~\ref{lem:integral-optimal-solutions} in Appendix~\ref{apdx:integralpf}. We mention here that Lemma~\ref{lem:integral-optimal-solutions} has an important implication on computing efficient (integral) assignments. To this end, note that the total number of decision variables and constraints of the linear Program~\eqref{eq:ShapleyShubikObjFull}-\eqref{eq:SSdistributionalUB-constraintsFull} are polynomial in the number of firms, workers, and the cardinality of the constraint set given by $\sum_{f \in F} |\H_f|$. As a result, if the upper bound hiring constraints are polymatroid constraints, then efficient assignments can be computed through Problem~\eqref{eq:ShapleyShubikObjFull}-\eqref{eq:SSdistributionalUB-constraintsFull}, using methods such as in~\cite{KHACHIYAN198053}\footnote{Under polymatroid hiring constraints, all the vertices of the polyhedron corresponding Constraints~\eqref{eq:SSWorkerAllocation}-\eqref{eq:SSdistributionalUB-constraintsFull} are integral~\cite{TAKAZAWA20192002}. Thus, any polynomial time algorithm for linear programs, e.g., Karmarkar's algorithm~\cite{karmarkar-1984} or variants of the ellipsoid method by Khachiyan~\cite{KHACHIYAN198053}, that finds basic optimal solutions, i.e., vertices of the polyhedron, can be used to compute efficient (integral) assignments. Note that the simplex algorithm will likely achieve fast practical performance in computing efficient assignments, even though it is not guaranteed to converge to the optimal solution in polynomial time.}, in polynomial time in the number of firms, workers, and the cardinality of the constraint set.

\subsection{Computing Salaries using Duality} \label{sec:mainExistence}

In this section, we complete the proof of Theorem~\ref{thm:ExistenceHierarchy} by using linear programming duality to establish the existence of salaries $\{\s_f \}_{f \in F}$ such that the arrangement $(\Xb^*, \{\s_f \}_{f \in F})$, where $\Xb^*$ is an efficient assignment, is stable under the condition that the linear Program~\eqref{eq:ShapleyShubikObjFull}-\eqref{eq:SSdistributionalUB-constraintsFull} has an integral optimal solution.

We first state the main result of this section, which establishes that if there exists an integral optimal solution $\Xb^*$ to the linear Program~\eqref{eq:ShapleyShubikObjFull}-\eqref{eq:SSdistributionalUB-constraintsFull}, then there exist salaries $\{\s_f\}_{f \in F}$ such that the arrangement $(\Xb^*, \{\s_f\}_{f \in F})$ satisfies the necessary and sufficient condition for stability in Lemma~\ref{lem:NSC-Stability}.

\begin{lemma} [Duality and Stable Arrangements] \label{lem:NSC-Satisfied}
Suppose that there exists an optimal solution $\Xb^*$ to the linear Program~\eqref{eq:ShapleyShubikObjFull}-\eqref{eq:SSdistributionalUB-constraintsFull} such that $x_{w,f}^* \in \{ 0, 1 \}$ for all worker-firm pairs $(w,f)$. Then, there exist salaries $\{ \s_f \}_{f \in F}$ corresponding to payoff vectors $\u, \v$ for the workers and firms, respectively, such that the arrangement $(\Xb^*, \{ \s_f \}_{f \in F})$ is feasible, individually rational for workers and firms, and $\sum_{w \in D} \alpha_{w,f} \leq \sum_{w \in D} u_w + v_f$ for all feasible sets of workers $D \in \T_f \cup \emptyset$ for each firm $f \in F$.
\end{lemma}

\begin{proof}
To prove this claim, we use linear programming duality. In particular, we consider the dual of the linear Program~\eqref{eq:ShapleyShubikObjFull}-\eqref{eq:SSdistributionalUB-constraintsFull}, and use the existence of the solution to the dual problem to construct the payoff vectors $\u \in \mathbb{R}^{|W|}$ and $\v \in \mathbb{R}^{|F|}$ that satisfy the necessary and sufficient condition for stability established in Lemma~\ref{lem:NSC-Stability}. Observe that the existence of such payoffs implies the existence of salaries $\{ \s_f \}_{f \in F}$ such that the arrangement $(\Xb^*, \{ \s_f \}_{f \in F})$ is stable.

To define the dual of Problem~\eqref{eq:ShapleyShubikObjFull}-\eqref{eq:SSdistributionalUB-constraintsFull}, we introduce the dual variables $\mmu = ( \mu_w )_{w \in W}$ of the allocation Constraints~\eqref{eq:SSWorkerAllocation} and the dual variables $\Bar{\eeta} = \{ \Bar{\eta}_{D} \}_{D \in \H_f, f \in F}$ of the upper bound hiring Constraints~\eqref{eq:SSdistributionalUB-constraintsFull}. Then, the dual of the linear Program~\eqref{eq:ShapleyShubikObjFull}-\eqref{eq:SSdistributionalUB-constraintsFull} is given by
\begin{mini!}|s|[2]<b>
	{\mmu, \Bar{\eeta}}{\sum_{w \in W} \mu_w + \sum_{f \in F} \sum_{D \in \H_f} \Bar{\eta}_{D} \Bar{\lambda}_{D}, \label{eq:DualObjUB}} 
	{}
	{}
	\addConstraint{\alpha_{w,f} - \mu_w - \sum_{D \in \H_f} \mathbbm{1}_{w \in D} \Bar{\eta}_{D} }{\leq 0, \quad \forall w \in W, f \in F \label{eq:stabilityConstraintUB},}
	\addConstraint{ \mmu}{ \geq \0, \quad \Bar{\eeta} \geq \0 \label{eq:signConstraintsUB}.}
\end{mini!}
Here, we let $\mmu^*$ and $\Bar{\eeta}^*$ be an optimal solution to the dual Problem~\eqref{eq:DualObjUB}-\eqref{eq:signConstraintsUB}.

We now establish the existence of salaries $\{ \s_f \}_{f \in F}$ such that the arrangement $(\Xb^*, \{ \s_f \}_{f \in F})$ is stable by constructing payoff vectors $\u \in \mathbb{R}^{|W|}$ and $\v \in \mathbb{R}^{|F|}$ that satisfy the necessary and sufficient condition for stability in Lemma~\ref{lem:NSC-Stability}. To this end, first observe by the existence of an integral optimal solution $\Xb^*$ to the primal linear program that a feasible solution to the dual Problem~\eqref{eq:DualObjUB}-\eqref{eq:signConstraintsUB} exists and that strong duality holds. Next, we construct worker and firm payoffs based on the optimal solution of the dual Problem~\eqref{eq:DualObjUB}-\eqref{eq:signConstraintsUB} such that the payoffs of the workers is given by $u_w = \mu_w^*$ for all $w \in W$ and that of the firms be given by $v_f = \sum_{D \in \H_f} \Bar{\eta}_{D}^* \Bar{\lambda}_{D}$ for all firms $f \in F$. Note that these payoffs can be realized under the salaries where $s_{w,f} = u_w - a_{w,f}$ for all worker-firm pairs $(w,f)$ with $x_{w,f}^* = 1$, and $s_{w,f} < - \max_{(w,f) \in W \times F} a_{w,f}$ otherwise. We now show that the arrangement $(\Xb^*, \{ \s_f \}_{f \in F})$ is stable by leveraging Lemma~\ref{lem:NSC-Stability}.

To see this, first observe that the arrangement is feasible since the assignment $\Xb^*$ is feasible, and the arrangement is individually rational for both the workers and firms since the payoffs are non-negative by the sign Constraints~\eqref{eq:signConstraintsUB}, i.e., $u_w = \mu_w^* \geq 0$ for all workers $w \in W$ and $v_f = \sum_{D \in \H_f} \Bar{\eta}_{D}^* \Bar{\lambda}_{D} \geq 0$ for all firms $f \in F$.

Finally, we establish that the payoff vectors $\u,\v$ satisfy the necessary and sufficient condition for stability in Lemma~\ref{lem:NSC-Stability} by showing that for all feasible sets of workers $D$ for each firm $f$ that $\sum_{w \in D} u_w + v_f \geq b_{D,f} + \sum_{w \in D} a_{w,f} = \sum_{w \in D} \alpha_{w,f}$. To establish this claim, first note that if the set $D = \emptyset$, then this result trivially holds by the sign Constraints~\eqref{eq:signConstraintsUB}. Next, fix a firm $f$ and a set of feasible workers $D \in \T_f$. Then, we know by the feasibility Constraints~\eqref{eq:stabilityConstraintUB} of the dual problem that $\alpha_{w,f} \leq \mu_w^* + \sum_{D \in \H_f} \mathbbm{1}_{w \in D} \Bar{\eta}_{D}^*$ for all worker-firm pairs $(w, f)$. Summing this relation over all workers $w \in D$, we get that
\begin{align*}
    \sum_{w \in D} \alpha_{w,f} &\leq 
    \sum_{w \in D} \mu_w^* + \sum_{w \in D} \sum_{D \in \H_f} \mathbbm{1}_{w \in D} \Bar{\eta}_{D}^* \stackrel{(a)}{\leq} \sum_{w \in D} \mu_w^* + \sum_{D \in \H_f} \Bar{\eta}_{D}^* \Bar{\lambda}_{D} \stackrel{(b)}{=} \sum_{w \in D} u_w + v_f,
\end{align*}
where (a) follows by sign Constraints~\eqref{eq:signConstraintsUB} and the fact that for any feasible set of workers $D \in \T_f$ it holds that $\sum_{w\in D} \mathbbm{1}_{w \in D} \leq \Bar{\lambda}_{D}$, 
and (b) follows by the above definitions of the worker and firm payoffs, $\u$ and $\v$, respectively.

Since $D$ is an arbitrary feasible set for firm $f$, the above inequality $\sum_{w \in D} u_w + v_f \geq \sum_{w \in D} \alpha_{w,f}$ holds for all feasible sets $D \in \T_f \cup \emptyset$ for any firm $f \in F$. Thus, we have shown that the arrangement $(\Xb^*, \{\s_f \}_{f \in F})$ satisfies the sufficient condition for stability in Lemma~\ref{lem:NSC-Stability}, which proves our claim.
\end{proof}
The proof of Lemma~\ref{lem:NSC-Satisfied} implies that any solution to the dual linear Program~\eqref{eq:DualObjUB}-\eqref{eq:signConstraintsUB} corresponds to a vector of payoffs associated with a stable arrangement as long as the linear program has an integral optimal solution. However, we mention that there may be stable arrangements that are not solutions to the dual linear program and refer to Appendix~\ref{apdx:nonUniqueness} for an example of a market instance when this is the case. We note that this is in contrast to the one-to-one matching setting considered by Shapley and Shubik~\cite{shapley1971assignment}, wherein the set of all stable arrangements are exactly the set of all solutions of the corresponding dual linear program. Furthermore, we mention that the structure of the payoffs of the workers and firms under stable arrangements in the many-to-one matching setting under distributional constraints also differs from the corresponding payoff structure in the one-to-one matching setting of Shapley and Shubik~\cite{shapley1971assignment} (see Appendix~\ref{apdx:nonUniqueness} for an example). Thus, we observe that the addition of distributional constraints fundamentally alters the properties of stable arrangements.

Observe that jointly Lemmas~\ref{lem:integral-optimal-solutions} and~\ref{lem:NSC-Satisfied} establish Theorem~\ref{thm:ExistenceHierarchy}. Since both the efficient assignment $\Xb^*$ and the salaries $\{\s_f \}_{f \in F}$, under polymatroid constraints on hiring, can be computed through the linear Program~\eqref{eq:ShapleyShubikObjFull}-\eqref{eq:SSdistributionalUB-constraintsFull} and its dual, it follows that the stable arrangement $(\Xb^*, \{\s_f \}_{f \in F})$ can be computed in polynomial time in the number of firms, workers, and the cardinality of the constraint set.

\subsection{Generalization to Lower Bound Constraints} \label{sec:lowerBoundGeneral}

In this section, we present the generalization of Theorem~\ref{thm:ExistenceHierarchy} to the setting where we additionally consider lower bound hiring constraints. In this setting, note that in contrast to upper bound hiring constraints, where remaining unmatched is feasible and individually rational for all firms, assignments that respect the lower bound hiring constraints may not correspond to individually rational outcomes for firms. In particular, the individual rationality of firms and the feasibility of an assignment under the lower bound hiring constraints may be at odds (see Appendix~\ref{apdx:IRodds} for an example), which may preclude the existence of stable arrangements. As a result, to generalize Theorem~\ref{thm:ExistenceHierarchy} to the setting of lower bound constraints, we consider a new notion of stability where we do not require the individual rationality of firms, as in Kojima et al.~\cite{jobMatchingConstraints2020}. Under this new notion of stability, we present a sufficient condition on the distributional constraints for stable arrangements to exist and show that such arrangements can be efficiently computed using linear programming.

We first present the new, relaxed, notion of stability, which we term r-stability, wherein we relax the individual rationality requirement of firms and instead require that each firm $f$ is assigned to a subset of workers in their feasibility collection $\T_f$. To this end, we first define an assignment $\Xb$ to be \emph{r-feasible} if the set of workers $D$ assigned to each firm $f$ satisfy its distributional constraints, i.e., $D \in \T_f$. Note that under r-feasibility, firms cannot remain unmatched unless $\emptyset \in \T_f$. Thus, we interpret this feasibility condition as the firms receiving a payoff of $-\infty$ if the assignment is infeasible, which thus makes it ``individually rational'' for the firms to obtain a r-feasible assignment even if it results in negative payoffs. We now present the notion of an \emph{r-stable} arrangement, as in~\cite{jobMatchingConstraints2020}. 

\begin{definition} [r-Stable Arrangement] \label{def:r-stability}
An arrangement $(\Xb, \{ \s_f \}_{f \in F})$ is r-stable if it is individually rational for all workers, the assignment $\Xb$ is r-feasible, and there is no firm-worker set $ \{ f \} \cup D$, where $D$ is r-feasible for firm $f$, i.e., $D \in \T_f$, and vector of salaries $\r_f$, such that
\begin{align*}
    &a_{w,f} + r_{w,f} \geq a_{w,f_w} + s_{w,f_w}, \quad \text{for all } w \in D, \text{ and } \\
    &b_{D, f} - \sum_{w \in D} s_{w,f} \geq b_{D_f, f} - \sum_{w \in D_f} s_{w,f},
\end{align*}
hold, with a strict inequality holding for at least one member in $D \cup \{ f \}$. Here each worker $w$ is assigned to firm $f_w$, and a set of workers $D_f$ is assigned to each firm $f$ under the assignment $\Xb$.
\end{definition}
We mention here that our earlier results obtained under the notion of stability presented in Definition~\ref{def:stability} naturally generalize under the notion of r-stability. In particular, we present the generalizations of Lemma~\ref{lem:NSC-Stability}, Theorem~\ref{thm:stableOptimal}, and Theorem~\ref{thm:ExistenceOneFirm} in Appendix~\ref{apdx:r-stability-extensions}.

We now present the main result of this section, which establishes a connection between linear programming duality and the existence of stable arrangements under a condition on the constraint structure and hiring quotas that correspond to \emph{generalized polymatroid constraints} (see Appendix~\ref{apdx:generalizedDefs} for a definition).

\begin{theorem} [Linear Programming and r-Stable Arrangements] \label{thm:ExistenceHierarchy2}
Suppose that for each firm $f$ the constraint structure $\H_f \times \{f\}$ is an intersecting family on $W \times F$, and there are integer-valued functions $g_f$, $\rho_f$ on $\H_f \times \{f \}$, that are submodular and supermodular, respectively, on intersecting sets such that $\rho_f(D) - \rho_f(D \backslash D') \leq g_f(D') - g_f(D' \backslash D)$ for all $D, D' \in \H_f$ and firms $f \in F$. Further, suppose that the upper and lower bound hiring quotas $\Bar{\llambda}_f, \underline{\llambda}_f$ satisfy $\Bar{\lambda}_{D} = g_f(D \times \{f\})$ and $\underline{\lambda}_{D} = \rho_f(D \times \{f\})$ for all $D \in \H_f$ and firms $f \in F$. Then, there exists a r-stable arrangement and this arrangement can be computed, using linear programming, in polynomial time in the number of firms, workers, and the cardinality of the constraint set given by $\sum_{f \in F} |\H_f|$.
\end{theorem}
The proof of Theorem~\ref{thm:ExistenceHierarchy2} is analogous to that of Theorem~\ref{thm:ExistenceHierarchy}, and is presented in Appendix~\ref{apdx:pfLB}. We note here that the given condition on the constraint structure presented in the statement of Theorem~\ref{thm:ExistenceHierarchy2} reduces to the condition in the statement of Theorem~\ref{thm:ExistenceHierarchy} representing polymatroid constraints if the supermodular functions $\rho_f$ capturing the lower bound constraints are identically zero and the submodular functions $g_f$ are non-decreasing, i.e., $g_f(A) \leq g_f(B)$ if $A \subseteq B$, for all firms $f$ . Furthermore, the condition on the constraint structure in Theorem~\ref{thm:ExistenceHierarchy2} closely resembles the necessary and sufficient condition on the set of constraints that preserve the substitutes condition, established in~\cite{jobMatchingConstraints2020}
\footnote{Kojima et al.~\cite{jobMatchingConstraints2020} consider \emph{generalized polyhedral constraints}, which are closely related to the conditions on the constraint structure in the statement of Theorem~\ref{thm:ExistenceHierarchy2}. For the ease of exposition, we focus on \emph{generalized polymatroid constraints} in Theorem~\ref{thm:ExistenceHierarchy2} and remark that this result extends even under \emph{generalized polyhedral constraints} (see Remark~\ref{rmk:gpc} in Appendix~\ref{apdx:pfLB}).}. Finally, we mention that this result builds on Theorem~\ref{thm:ExistenceHierarchy} by further generalizing the duality theory approach of Shapley and Shubik~\cite{shapley1971assignment} to an even broader class of constraints that include both upper and lower bound hiring constraints.

\section{Conclusion and Future Work} \label{sec:conclusion}

In this work, we studied two-sided many-to-one matching markets with transferable utilities subject to distributional constraints on the set of feasible allocations. In such markets, we established the efficiency of stable arrangements and, in the setting of one firm, showed their existence irrespective of the nature of the constraint structure or agent preferences. For markets with multiple firms, we showed that, even when firms have linear preferences over workers, stable arrangements may not exist under general constraint structures. Thus, for markets with linear preferences of firms, we investigated sufficient conditions on the constraint structure to guarantee the existence of stable arrangements and developed a method compute such arrangements efficiently. To this end, we developed an approach based on linear programming duality that extended Shapley and Shubik's~\cite{shapley1971assignment} duality theory approach in the one-to-one matching setting to the many-to-one matching setting under distributional constraints. This duality-based approach provided a method to compute stable arrangements efficiently through a connection between linear programming duality and stability.


There are various directions for future research. First, it would be worthwhile to study whether the linear programming approach can guarantee the existence of stable arrangements for constraints beyond polymatroid and generalized polymatroid constraints, i.e., for constraint structures that do not satisfy the substitutes condition. Next, it would be interesting to investigate whether the linear programming duality approach can be generalized to a broader range of firm preferences through methods such as convex programming. Finally, in many-to-one matching markets with transfers and distributional constraints, it would also be valuable to explore mechanisms that, in addition to preserving stability, are strategy-proof for agents on one side of the market.

\bibliographystyle{unsrt}
\bibliography{main}

\appendix

\section{Additional Theoretical Analyses and Results}

\subsection{Proof of Lemma~\ref{lem:NSC-Stability}} \label{apdx:pfnsc-stability}

One direction of this claim follows directly from the definition of stability. To see this, first note that by definition a stable arrangement is both feasible and individually rational for workers and firms. Furthermore, observe for a firm $f$ and any feasible set of workers $D$ that the cumulative payoff received by the firm-worker set $\{ f\} \cup D$ is $\sum_{w \in D} a_{w,f} + b_{D,f}$ if matched. Next, if for some some firm $f$ and for some set of workers $D \in \T_f \cup \emptyset$, it holds under the arrangement $(\Xb^*, \{ \s_f \}_{f \in F})$ that
\begin{align*}
    \sum_{w \in D} a_{w,f} + b_{D,f} > \sum_{w \in D} u_w + v_f,
\end{align*}
then the firm $f$ and the workers in the set $D$ can profitably deviate by matching with each other and obtaining a cumulative match value of at least $\sum_{w \in D} a_{w,f} + b_{D,f}$. Since $(\Xb^*, \{ \s_f \}_{f \in F})$ is stable, such a deviation by a coalition of workers and a firm is not possible, thereby proving one direction of this claim.

To prove the other direction, suppose that an arrangement $(\Xb^*, \{ \s_f \}_{f \in F})$ is feasible, individually rational for workers and firms, and for all feasible sets of workers $D$ for all firms $f \in F$ it holds that $\sum_{w \in D} a_{w,f} + b_{D,f} \leq \sum_{w \in D} u_w + v_f$. Then, we claim that there is no feasible set $D \in \T_f \cup \emptyset$ of workers for a given firm $f$ such that the firm $f$ as well as all the workers in that set $D$ are weakly better off with at least one worker in the set $D$ or the firm $f$ being strictly better off. Suppose for contradiction that such a coalition of workers $D' \in \T_f \cup \emptyset$ and a firm $f$ exists that can profitably deviate with at least one of the members of the coalition being strictly better off. To this end, let $\Tilde{u}_w$ for $w \in D'$ and $\Tilde{v}_f$ denote the payoffs for agents in the firm-worker set $\{ f \} \cup D'$ when the workers in set $D'$ match with firm $f$. Then, the cumulative payoff received by this firm-worker set $\{ f \} \cup D'$ is $\sum_{w \in D'} a_{w,f} + b_{D',f}$. Note that since this was a profitable deviation for this coalition, it must hold that $\Tilde{u}_w \geq u_w$ for all $w \in D'$ and $\Tilde{v}_f \geq v_f$ with at least one of the inequalities being strict. As a result, we get that
\begin{align*}
    \sum_{w \in D'} a_{w,f} + b_{D',f} \stackrel{(a)}{=} \sum_{w \in D'} \Tilde{u}_{w} + \Tilde{v}_{f} \stackrel{(b)}{>} \sum_{w \in D'} u_{w} + v_{f} \stackrel{(c)}{\geq}\sum_{w \in D'} a_{w,f} + b_{D',f},
\end{align*}
a contradiction. Here (a) follows from the fact that the cumulative payoffs of agents in the firm-worker set $\{ f \} \cup D'$ is $\sum_{w \in D'} a_{w,f} + b_{D,f}$ when workers in the set $D'$ match with the firm $f$, (b) follows from the fact that the coalition $\{ f \} \cup D'$ had a profitable deviation, and (c) holds by the assumption on the arrangement $(\Xb^*, \{ \s_f \}_{f \in F})$. Since we have obtained our desired contradiction, we have shown the second direction of our claim.
\qed

\subsection{Proof of Lemma~\ref{lem:integral-optimal-solutions}} \label{apdx:integralpf}

To prove this result, we leverage the polymatroid intersection theorem, given in Theorem~\ref{thm:polyintthm}. In particular, we show under the condition on the upper bound hiring constraints in the statement of the lemma that the linear Program~\eqref{eq:ShapleyShubikObjFull}-\eqref{eq:SSdistributionalUB-constraintsFull} is a special case of the linear program $\max_{\y \in \mathbb{R}^{|E|}} \c^T \y \text{ s.t. } \y \in P_1 \cap P_2$ for two polymatroids $P_1, P_2$, a set $E$ and a vector $\q$.

To this end, we define the set $E = W \times F$, i.e., the set of all tuples of worker-firm pairs $(w,f)$, the objective coefficients $\q$ as the vector with entries $\alpha_{w,f}$ and the decision variables $\y = (y_j)_{j \in E}$ as a vector $\x$ with entries $x_{w,f}$. Further, the polyhedron
\begin{align*}
    P_f = \left\{ \x: \x \geq \0, \sum_{ \genfrac{}{}{0pt}{2}{(w, f) \in}{D \times \{ f \}} } x_{w,f} \leq g_f(D \times \{f\}), \text{ for } D \in \H_f \right\}
\end{align*}
for each firm $f$ is a polymatroid, since by assumption $\H_f \times \{f\}$ is an intersecting family and the function $g_f$ on $\H_f \times \{f\}$ is a submodular on intersecting sets.

Next, we note that $P_F = \cap_{f \in F} P_f$ is also a polymatroid. To see this, note first that each $\H_f \times \{f\}$ is intersecting. Next, noting that for any $A \in \H_f \times \{f\}$, $B \in \H_{f'} \times \{f'\}$ for $f \neq f'$ that $A \cap B = \emptyset$, it holds that $\cup_{f \in F} \H_f \times \{f\}$ is intersecting. Thus, it follows that $P_F = \cap_{f \in F} P_f$ is a polymatroid.

We now show that the allocation Constraints~\eqref{eq:SSWorkerAllocation} also correspond to polymatroid constraints. To this end, define a family $\H_W$ for the workers given by the sets $\{ w \} \times F$ for all workers $w \in W$. Note clearly that the family $\H_W$ is intersecting, since for any $A_1, A_2 \in \H_W$, such that $A_1 \neq A_2$ it holds that $A_1 \cap A_2 = \emptyset$. Thus, letting the set function $g(\{w\} \times F) = 1$ for all $w \in W$, we have that
\begin{align*}
    P_W = \{ \x: \x \geq \0, \sum_{ \genfrac{}{}{0pt}{2}{(w,f) \in}{\{ w \} \times F}} x_{w,f} \leq 1, \text{ for } w \in W \}
\end{align*}
is a polymatroid.

Thus, we have that the linear Program~\eqref{eq:ShapleyShubikObjFull}-\eqref{eq:SSdistributionalUB-constraintsFull} can be rewritten as
$$\max_{\x \in \mathbb{R}^{|W \times F|}} \sum_{(w,f) \in W \times F} \alpha_{w,f} x_{w,f} \text{ s.t. } \x \in P_W \cap P_F.$$

Then, from the polymatroid intersection theorem (Theorem~\ref{thm:polyintthm}), it follows that the linear Program~\eqref{eq:ShapleyShubikObjFull}-\eqref{eq:SSdistributionalUB-constraintsFull} has an integral optimal solution.

\subsection{Extensions of Results under r-Stability} \label{apdx:r-stability-extensions}

In this section, we present the natural extensions of Lemma~\ref{lem:NSC-Stability}, Theorem~\ref{thm:stableOptimal}, and Theorem~\ref{thm:ExistenceOneFirm} under the notion of r-stability. We omit the proofs of these extensions since their proofs are almost entirely analogous to the proofs of their counterparts under the stability notion of Definition~\ref{def:stability}. We mention here that the primary difference in the proofs of the extensions of these results is that under the notion of r-stability, any firm $f$ must hire a set of workers $D$ that satisfy the hiring constraints, i.e., $D \in \T_f$. Note that this is in contrast to the stability notion of Definition~\ref{def:stability}, wherein firms had the option of remaining unmatched, i.e., any firm $f$ could hire a set of workers $D \in \T_f \cup \emptyset$.

\paragraph{Necessary and Sufficient Condition for r-Stability} 

We first present the necessary and sufficient condition for r-stability, which generalizes Lemma~\ref{lem:NSC-Stability}.

\begin{lemma} [Necessary and Sufficient Condition for r-Stability] \label{lem:NSC-r-Stability}
Let $\u = (u_w)_{w \in W}$ and $\v = (v_f)_{f \in F}$ represent the payoff vectors of the workers and the firms under the arrangement $(\Xb^*, \{ \s_f \}_{f \in F})$. Then, an arrangement $(\Xb^*, \{ \s_f \}_{f \in F})$ is r-stable if and only if it is r-feasible, individually rational for workers, and it holds for each r-feasible set of workers $D \in \T_f$ for each firm $f \in F$ that $\sum_{w \in D} a_{w,f} + b_{D,f} \leq \sum_{w \in D} u_w + v_f$, i.e., the sum of the payoffs of a firm $f$ and any r-feasible set of workers $D$ is at least the sum of their match values $\sum_{w \in D} a_{w,f} + b_{D,f}$.
\end{lemma}

Note that the above condition is analogous to the necessary and sufficient condition for stability in Lemma~\ref{lem:NSC-Stability} other than the requirement of r-feasibility and the relaxation of the individual rationality of firms.

\paragraph{Efficiency of r-Stable Arrangements}

To present the extension of Theorem~\ref{thm:stableOptimal} under this new notion of stability, we first define an appropriate notion of the efficiency of assignments in this context, which we refer to as r-efficiency.

\begin{definition} [r-Efficiency of Assignments] \label{def:r-efficiency}
We define an assignment $\Xb^*$ of workers to firms to be \emph{r-efficient} if it maximizes the total match value among the class of all r-feasible assignments, i.e., $U(\Xb^*) \geq U(\Xb)$ for all r-feasible assignments $\Xb$.
\end{definition}
Note that, as opposed to the efficiency of assignments, the notion of r-efficiency is defined with respect to the set of all r-feasible assignments, i.e., the assignment must be feasible with respect to the firm's hiring constraints and thus firm $f$ cannot remain unmatched unless $\emptyset \in \T_f$. We now present the efficiency result for r-stable arrangements.

\begin{theorem} [r-Efficiency of r-Stable Arrangements] \label{thm:r-stableOptimal}
If an arrangement $(\Xb^*, \{ \s_f \}_{f \in F})$ is r-stable, then the assignment $\Xb^*$ is r-efficient.
\end{theorem}

\paragraph{Existence of r-Stable Arrangements in a One-Firm Setting}

Finally, we present the extension of Theorem~\ref{thm:ExistenceOneFirm}, which, under the notion of stability in Definition~\ref{def:stability}, establishes the existence of stable arrangements under general agent preferences and constraints, under r-stability. 


\begin{theorem} [Existence of r-Stable Arrangement for One Firm] \label{thm:r-ExistenceOneFirm}
In a one-firm setting, there exists a r-stable arrangement under any vector of valuations $\b = (b_{D,f})_{D \in \T_f}$ of the firm $f$ and any feasibility collection $\T_f$ for which there exists a r-feasible assignment.
\end{theorem}
We mention that, as opposed to Theorem~\ref{thm:ExistenceOneFirm}, we require here that an r-feasible assignment exists, as otherwise no r-stable arrangement can form. Note that no such feasibility requirement was needed for Theorem~\ref{thm:ExistenceOneFirm} since under the notion of stability in Definition~\ref{def:stability} it was feasible for firms to remain unmatched.

\subsection{Generalized Polymatroid Constraints} \label{apdx:generalizedDefs}

Here, we introduce the notion of generalized polymatroid constraints, as in Frank~\cite{frank1984generalized}.

\begin{definition} [Generalized Polymatroid Constraints] \label{def:g-polymatroids}
Let $\H'$ be an intersecting family over a set S and $g,\rho$ be integer-valued functions on $\H'$ that are submodular and supermodular, respectively, on intersecting sets where $\rho(D) - \rho(D \backslash D') \leq g(D') - g(D' \backslash D)$ for all $D, D' \in \H'$. Then, the polyhedron
\begin{align*}
    P = \{ \y: \y \geq \0, \rho(D) \leq \sum_{d \in D} y_d \leq g(D), \text{ for } D \in \H' \}
\end{align*}
defines a generalized polymatroid $P(\H', g, \rho)$, and the corresponding constraints are generalized polymatroid constraints.
\end{definition}

\subsection{Proof of Theorem~\ref{thm:ExistenceHierarchy2}} \label{apdx:pfLB}

\paragraph{Proof Outline}

To prove Theorem~\ref{thm:ExistenceHierarchy2}, we use Lemma~\ref{lem:NSC-r-Stability} and Theorem~\ref{thm:r-stableOptimal} to show that there exists some r-stable arrangement $(\Xb^*, \{s_f \}_{f \in F})$ that can be computed via a linear program that has a number of decision variables and constraints that are polynomial in the number of firms, workers, and the cardinality of the constraint set given by $\sum_{f \in F} |\H_f|$. To compute the assignment $\Xb^*$, we note by the r-efficiency of r-stable arrangements (Theorem~\ref{thm:r-stableOptimal}) that the assignment must maximize the total match value. Thus, we formulate a linear program analogous to Problem~\eqref{eq:ShapleyShubikObjFull}-\eqref{eq:SSdistributionalUB-constraintsFull} that also includes lower bound constraints on hiring to compute r-efficient fractional assignments. We then show that this linear program has an integral optimal solution under the specified condition on the constraint structure in the statement of the theorem. We mention that the number of decision variables and constraints of this linear program are polynomial in the number of firms, workers, and the cardinality of the constraint set. Then, we use linear programming duality to establish that there exist a vector of salaries $\{\s_f \}_{f \in F}$ such that the resulting payoffs of agents under the assignment $\Xb^*$ and salaries $\{\s_f \}_{f \in F}$ satisfy the necessary and sufficient condition for r-stability (Lemma~\ref{lem:NSC-r-Stability}), which proves our claim.

In the remainder of this section, we present the details of the above proof sketch.

\paragraph{Linear Programming to Compute r-Efficient Assignments} Since r-stable arrangements are r-efficient, we now present the following linear program with both upper and lower bound hiring constraints to compute r-efficient fractional assignments

\begin{maxi!}|s|[2]<b>
	{\Xb \in \mathbb{R}^{|W| \times |F|}}{U(\Xb) =\sum_{w \in W} \sum_{f \in F} \alpha_{w,f} x_{w,f}, \label{eq:ShapleyShubikObjFullLB}}
	{}
	{}
	\addConstraint{\sum_{f \in F} x_{w,f}}{\leq 1, \quad \forall w \in W \label{eq:SSWorkerAllocationLB},}
	\addConstraint{x_{w,f}}{ \geq 0, \quad \forall w \in W, f \in F \label{eq:SSnonnegativity-constraintsFullLB},}
	\addConstraint{ \sum_{w \in D} x_{w,f}}{ \leq \Bar{\lambda}_{D}, \quad \forall D \in \H_f, f \in F \label{eq:SSdistributionalUB-constraintsFullLB},}
	\addConstraint{ \sum_{w \in D} x_{w,f}}{ \geq \Bar{\lambda}_{D}, \quad \forall D \in \H_f, f \in F \label{eq:SSdistributionalLB-constraintsFullLB},}
\end{maxi!}
Note that the linear Program is identical to Problem~\eqref{eq:ShapleyShubikObjFull}-\eqref{eq:SSdistributionalUB-constraintsFull} other than that it additionally includes lower bound hiring Constraints~\eqref{eq:SSdistributionalLB-constraintsFullLB}. 

We now show that the solution to the linear Program~\eqref{eq:ShapleyShubikObjFullLB}-\eqref{eq:SSdistributionalLB-constraintsFullLB} is integral when the upper and lower bound hiring constraints correspond to generalized polymatroid constraints (Definition~\ref{def:g-polymatroids}).

\begin{lemma} [Integral Optimal Solutions With Lower Bound Constraints] \label{lem:r-integral-optimal-solutions}
Suppose that for each firm $f$ the constraint structure $\H_f \times \{f\}$ is an intersecting family over $W \times F$, and there are integer-valued functions $g_f$, $\rho_f$ on $\H_f \times \{f \}$, that are submodular and supermodular, respectively, on intersecting sets such that $\rho_f(D) - \rho_f(D \backslash D') \leq g_f(D') - g_f(D' \backslash D)$ for all $D, D' \in \H_f$ and firms $f \in F$. Further, suppose that the upper and lower bound hiring quotas $\Bar{\llambda}_f, \underline{\llambda}_f$ satisfy $\Bar{\lambda}_{D} = g_f(D \times \{f\})$ and $\underline{\lambda}_{D} = \rho_f(D \times \{f\})$ for all $D \in \H_f$ and firms $f \in F$. Then there is an optimal solution $\Xb^*$ of the linear Program~\eqref{eq:ShapleyShubikObjFullLB}-\eqref{eq:SSdistributionalLB-constraintsFullLB} such that $x_{w,f}^* \in \{ 0, 1 \}$ for all worker-firm pairs $(w,f) \in W \times F$.
\end{lemma}
Note here that the condition on the constraint structure and hiring quotas in the statement of Lemma~\ref{lem:r-NSC-Satisfied} (and Theorem~\ref{thm:ExistenceHierarchy2}) correspond to generalized polymatroid constraints. The proof of Lemma~\ref{lem:r-integral-optimal-solutions} follows from the generalized polymatroid intersection theorem~\cite{frank1984generalized}, which states that the optimal solution of a linear program under the intersection of two generalized polymatroids is integral.

\begin{theorem} [Generalized Polymatroid Intersection Theorem~\cite{frank1984generalized}] \label{thm:polyintthmr}
For a set $E$, a vector $\q$, and two generalized polymatroids $P_1, P_2$, the linear program $\max_{\y \in \mathbb{R}^{|E|}} \q^T \y \text{ s.t. } \y \in P_1 \cap P_2$ has an integral optimal solution.
\end{theorem}
Using Theorem~\ref{thm:polyintthmr}, we can apply an analogous reasoning to that used in the proof of Lemma~\ref{lem:integral-optimal-solutions} to establish Lemma~\ref{lem:r-integral-optimal-solutions} and so we omit the details here. In the following, we note that the result of Lemma~\ref{lem:r-integral-optimal-solutions} also extends to the setting of generalized polyhedral constraints introduced by Kojima et al.~\cite{jobMatchingConstraints2020}.

\begin{remark} [Extension of Lemma~\ref{lem:r-integral-optimal-solutions} under Generalized Polyhedral Constraints] \label{rmk:gpc}
We mention that Lemma~\ref{lem:r-integral-optimal-solutions} extends to the setting of generalized polyhedral constraints, which are closely related to generalized polymatroid constraints. To this end, we note that generalized polyhedral constraints include not only upper and lower bound hiring constraints, with similar restrictions on the constraint structure as in the case of generalized polymatroid constraints, but also include ``always-hiring'' and ``never-hiring'' constraints. That is, in addition to the upper and lower bound hiring constraints given by $\H_f$ for each firm $f$ there are constraints that correspond to the setting when firms must hire a certain subset of workers $\Bar{E}_f$ or not hire a certain subset of workers $\underline{E}_f$. Here $\Bar{E}_f \cap \underline{E}_f = \emptyset$, and for all subsets of workers $D \in \H_f$, $D \cap (\Bar{E}_f \cup \underline{E}_f) = \emptyset$, for all firms $f$. We note that the constraint structure of the linear program that additionally consider these ``always'' and ``never'' hiring constraints can also be shown to be the intersection of two generalized polymatroids in a manner similar to that used in the proof of Lemma~\ref{lem:integral-optimal-solutions} due to the condition on the constraint structure for all firms $f$ and subsets of workers $D \in \H_f$ that $D \cap (\Bar{E}_f \cup \underline{E}_f) = \emptyset$.
\end{remark}

\paragraph{Computing Salaries using Duality}

Finally, to complete the proof of Theorem~\ref{thm:ExistenceHierarchy2}, we use linear programming duality to establish the existence of salaries $\{\s_f \}_{f \in F}$ such that the arrangement $(\Xb^*, \{\s_f \}_{f \in F})$ is r-stable under the condition that the linear Program~\eqref{eq:ShapleyShubikObjFullLB}-\eqref{eq:SSdistributionalLB-constraintsFullLB} has an integral optimal solution. To this end, we show that if there exists an integral optimal solution $\Xb^*$ to the linear Program~\eqref{eq:ShapleyShubikObjFullLB}-\eqref{eq:SSdistributionalLB-constraintsFullLB}, then there exist salaries $\{\s_f \}_{f \in F}$ such that the arrangement $(\Xb^*, \{\s_f \}_{f \in F})$ satisfies the necessary and sufficient condition for r-stability  in Lemma~\ref{lem:NSC-r-Stability}.

\begin{lemma} [Duality and r-Stable Arrangements] \label{lem:r-NSC-Satisfied}
Suppose that there exists an optimal solution $\Xb^*$ to the linear Program~\eqref{eq:ShapleyShubikObjFullLB}-\eqref{eq:SSdistributionalLB-constraintsFullLB} such that $x_{w,f}^* \in \{ 0, 1 \}$ for all worker-firm pairs $(w,f)$. Then, there exist salaries $\{ \s_f \}_{f \in F}$ corresponding to payoff vectors $\u, \v$ for the workers and firms, respectively, such that the arrangement $(\Xb^*, \{ \s_f \}_{f \in F})$ is r-feasible, individually rational for workers, and $\sum_{w \in D} \alpha_{w,f} \leq \sum_{w \in D} u_w + v_f$ for all feasible sets of workers $D \in \T_f$ for each firm $f \in F$.
\end{lemma}

\begin{proof}
The proof of this claim follows a similar argument to that of Lemma~\ref{lem:NSC-Satisfied} other than the fact that there are additional dual variables introduced with the addition of the lower bound constraints that need to be used to define the payoffs of the firms and workers. To this end, we first define the dual of Problem~\eqref{eq:ShapleyShubikObjFullLB}-\eqref{eq:SSdistributionalLB-constraintsFullLB} and then construct payoff vectors $\u, \v$ that satisfy the necessary and sufficient condition for r-stability in Lemma~\ref{lem:NSC-r-Stability}.

To define the dual of Problem~\eqref{eq:ShapleyShubikObjFullLB}-\eqref{eq:SSdistributionalLB-constraintsFullLB}, we introduce the dual variables $\mmu = ( \mu_w )_{w \in W}$ of the allocation Constraints~\eqref{eq:SSWorkerAllocationLB}, the dual variables $\Bar{\eeta} = \{ \Bar{\eta}_{D} \}_{D \in \H_f, f \in F}$ of the upper bound hiring Constraints~\eqref{eq:SSdistributionalUB-constraintsFullLB}, and the dual variables $\underline{\eeta} = \{ \underline{\eta}_{D} \}_{D \in \H_f, f \in F}$ of the lower bound hiring Constraints~\eqref{eq:SSdistributionalLB-constraintsFullLB}. Then, the dual of the linear Program~\eqref{eq:ShapleyShubikObjFullLB}-\eqref{eq:SSdistributionalLB-constraintsFullLB} is given by
\begin{mini!}|s|[2]<b>
	{\mmu, \Bar{\eeta}}{\sum_{w \in W} \mu_w + \sum_{f \in F}  \sum_{D \in \H_f} \left( \Bar{\eta}_{D} \Bar{\lambda}_{D} + \underline{\eta}_{D} \underline{\lambda}_{D} \right), \label{eq:DualObjLB}} 
	{}
	{}
	\addConstraint{\alpha_{w,f} - \mu_w - \sum_{D \in \H_f} \mathbbm{1}_{w \in D} \cdot \left( \Bar{\eta}_{D} + \underline{\eta}_{D} \right) }{\leq 0, \quad \forall w \in W, f \in F \label{eq:stabilityConstraintLB},}
	\addConstraint{ \mmu}{ \geq \0, \quad \Bar{\eeta} \geq \0, \quad \underline{\eeta} \leq \0. \label{eq:signConstraintsLB}}
\end{mini!}
Here we let $\mmu^*$, $\Bar{\eeta}^*$ and $\underline{\eeta}^*$ be the optimal solutions to the dual Problem~\eqref{eq:DualObjLB}-\eqref{eq:signConstraintsLB}. We now define the payoff vectors $\u, \v$ using the optimal dual solutions. In particular, let the payoffs of the workers $u_w = \mu_w^*$ for all $w \in W$ and that of the firms be given by $v_f = \sum_{D \in \H_f} \left( \Bar{\eta}_{D}^* \Bar{\lambda}_{D} + \underline{\eta}_{D}^* \underline{\lambda}_{D} \right)$ for all firms $f \in F$. Note that these payoffs can be realized under the salaries where $s_{w,f} = u_w - a_{w,f}$ for all worker-firm pairs $(w,f)$ such that $x_{w,f}^* = 1$, and $s_{w,f} < - \max_{(w,f) \in W \times F} a_{w,f}$ otherwise. We now show that the arrangement $(\Xb^*, \{ \s_f \}_{f \in F})$ is stable by leveraging Lemma~\ref{lem:NSC-r-Stability}.

To see this, first observe that the arrangement is r-feasible since the assignment $\Xb^*$ is r-feasible (it satisfies all the upper and lower bound hiring constraints), and the set of payoffs are individually rational for the workers since $u_w = \mu_w^* \geq 0$ for all workers $w \in W$ by the sign Constraints~\eqref{eq:signConstraintsLB}. Note that the individual rationality of the firms cannot be guaranteed since $\underline{\eeta}^* \leq \0$.

Finally, we establish that the payoff vectors $\u,\v$ satisfy the necessary and sufficient condition for stability in Lemma~\ref{lem:NSC-r-Stability} in a similar manner to that in the proof of Lemma~\ref{lem:NSC-Satisfied}. To this end, fix a firm $f$ and a set of workers $D \in \T_f$. Then, summing the dual feasibility Constraints~\eqref{eq:stabilityConstraintLB} over all workers $w \in D$, we get that
\begin{align*}
    \sum_{w \in D} \alpha_{w,f} &\leq 
    \sum_{w \in D} \mu_w^* + \sum_{w \in D} \sum_{D \in \H_f} \mathbbm{1}_{w \in D} \left( \Bar{\eta}_{D}^* + \underline{\eta}_{D}^* \right), \\ &\stackrel{(a)}{\leq} \sum_{w \in D} \mu_w^* + \sum_{D \in \H_f} \left( \Bar{\eta}_{D}^* \Bar{\lambda}_{D} + \underline{\eta}_{D}^* \underline{\lambda}_{D} \right), \\
    &\stackrel{(b)}{=} \sum_{w \in D} u_w + v_f,
\end{align*}
where (a) follows by the sign Constraints~\eqref{eq:signConstraintsLB} and the fact that for any feasible set of workers $D \in \T_f$ it holds that $\underline{}{\lambda}_{D} \leq \sum_{w\in D} \mathbbm{1}_{w \in D} \leq \Bar{\lambda}_{D}$, 
and (b) follows by the above definitions of the worker and firm payoffs, $\u$ and $\v$, respectively.

Since $D$ is an arbitrary r-feasible set of workers for firm $f$, we have that the above inequality $\sum_{w \in D} u_w + v_f \geq \sum_{w \in D} \alpha_{w,f}$ holds for all r-feasible sets $D \in \T_f$ for any firm $f \in F$. Thus, we have shown that the arrangement $(\Xb^*, \{\s_f \}_{f \in F})$ satisfies the necessary and sufficient condition for stability in Lemma~\ref{lem:NSC-Stability}, which proves our claim.
\end{proof}

Jointly, Lemmas~\ref{lem:r-integral-optimal-solutions} and~\ref{lem:r-NSC-Satisfied} establish Theorem~\ref{thm:ExistenceHierarchy2}. We also note that, as in the case of polymatroid constraints, under generalized polymatroid constraints all the vertices of the polyhedron corresponding Constraints~\eqref{eq:SSWorkerAllocationLB}-\eqref{eq:SSdistributionalLB-constraintsFullLB} are integral~\cite{TAKAZAWA20192002}. Thus, any polynomial time algorithm for linear programs, e.g., Karmarkar's algorithm~\cite{karmarkar-1984} or variants of the ellipsoid method by Khachiyan~\cite{KHACHIYAN198053}, that finds basic optimal solutions, i.e., vertices of the polyhedron, can be used to compute r-stable arrangements in polynomial time. Furthermore, as in the case of Cororllary~\ref{cor:ExistenceHierarchy}, a direct consequence of Theorem~\ref{thm:ExistenceHierarchy2} is that r-stable arrangements exist and can be computed efficiently for upper and lower bound constraints that correspond to a hierarchy. Finally, we mention that Theorem~\ref{thm:ExistenceHierarchy2} focused on the case when the upper and lower bound constraints were over the same set $\H_f$ of workers for the ease of exposition. We thus note that Theorem~\ref{thm:ExistenceHierarchy2} can also be readily extended to the setting when the set of workers $\H_f$ over which a firm has upper bound constraints are not necessarily the same as the set of workers over which the firm has lower bound constraints.

\section{Examples}

\subsection{Violation of Substitutes Condition} \label{apdx:substitutes-violation}

Here, we present a figure to represent the constraint structure, described in Example~\ref{eg:violationSubstitutes}, under which the substitutes condition is violated.

\begin{figure}[!ht]
      \centering
      \includegraphics[width=0.6\linewidth]{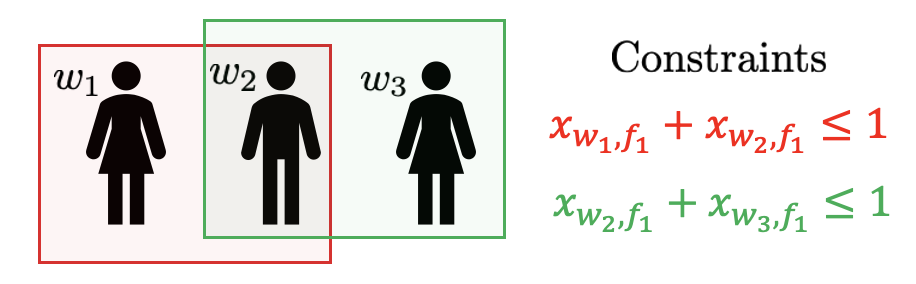}
      \caption{\small \sf Depiction of the constraint structure in a one-firm, three-worker example under which the substitutes condition is violated.}
      \label{fig:substitutes}
   \end{figure}

\subsection{Stable Arrangements May not Exist} \label{apdx:stablenonEx}

Here, we present a figure to represent the constraint structure, described in the proof of Proposition~\ref{prop:nonExistenceStable}, under which no stable arrangement exists.

\begin{figure}[!ht]
      \centering
      \includegraphics[width=0.85\linewidth]{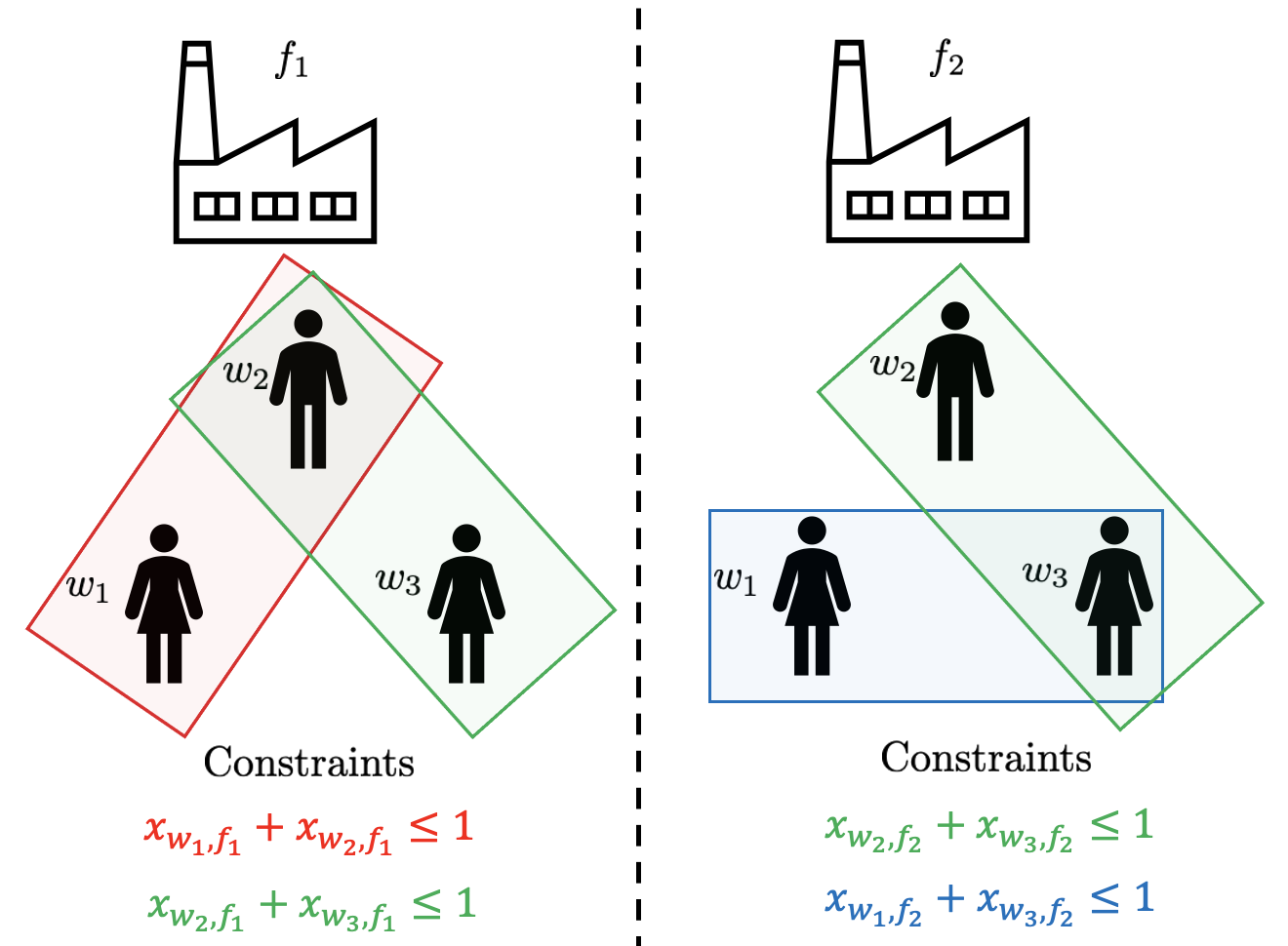}
      \caption{\small \sf Depiction of the constraint structure in a two-firm, three-worker example under which stable arrangements do not exist.}
      \label{fig:non-ex-stability}
   \end{figure}

\subsection{Polymatroid and Hierarchical Constraints} \label{eg:polyheirEg}

Here, we present a figure to represent the constraint structure, described in Example~\ref{eg:polyheirEg}, to elucidate the notions of hierarchical and polymatroid constraints. 

\begin{figure}[!ht]
      \centering
      \includegraphics[width=0.95\linewidth]{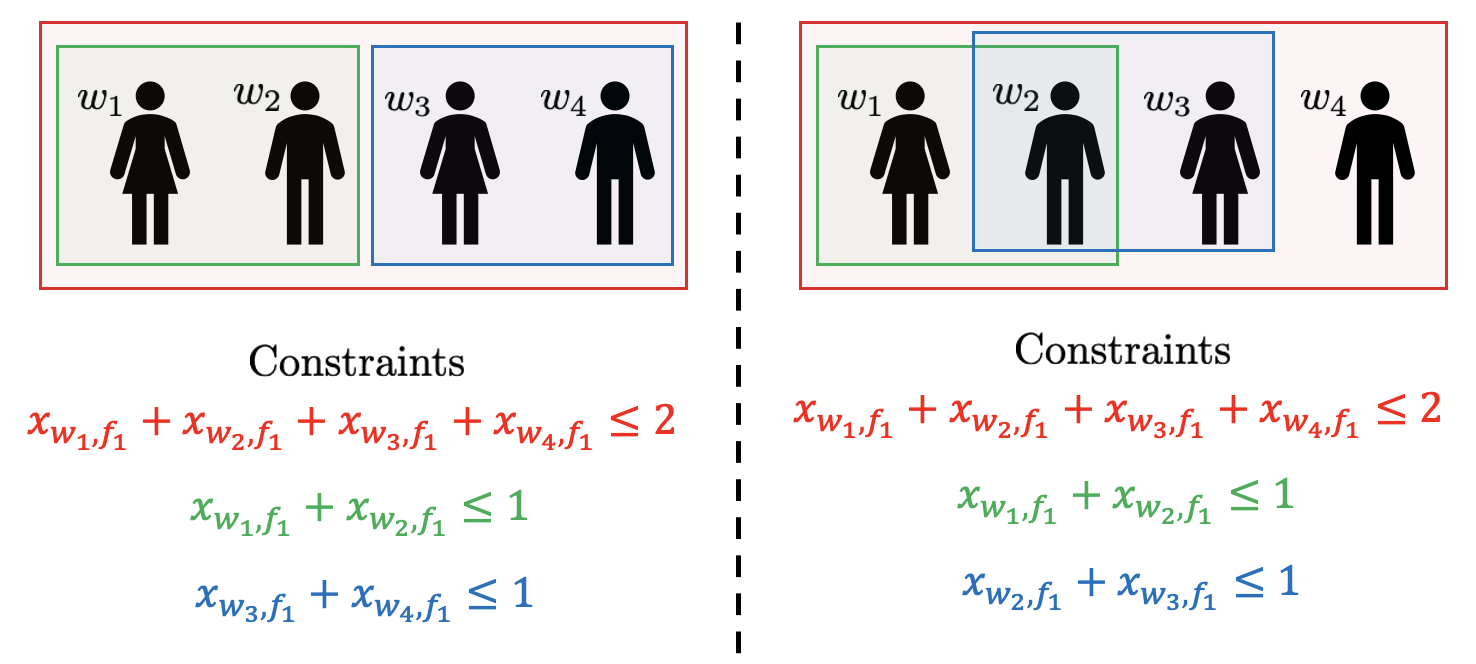}
      \caption{\small \sf Example depicting polymatroid and hierarchical constraints. On the left, the upper bound constraints correspond to a hierarchy since the hiring constraints are over sets of workers that are either disjoint or are subsets of each other. On the right, the constraints represent neither hierarchical nor polymatroid constraints.}
      \label{fig:polyheir}
   \end{figure}

\subsection{Non-Integrality of Solution of Linear Program} \label{apdx:egNonIntegrality}

\begin{figure}[!ht]
      \centering
      \includegraphics[width=0.6\linewidth]{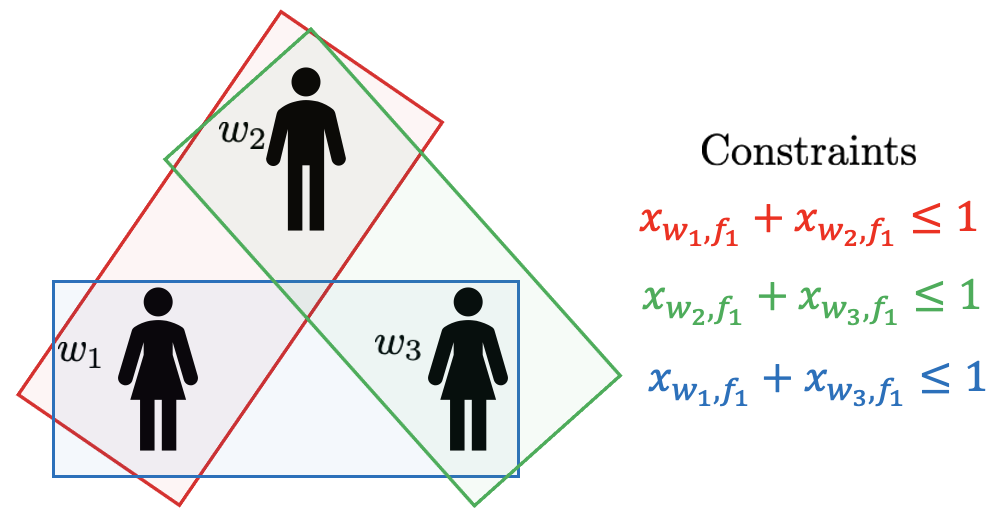}
      \caption{\small \sf Depiction of the constraint structure in a one-firm, three-worker example under which no integral solution to the linear Program~\eqref{eq:ShapleyShubikObjFull}-\eqref{eq:SSdistributionalUB-constraintsFull} exists if the match values of all worker-firm pairs are equal.}
      \label{fig:non-integrality}
   \end{figure}

We construct an instance with upper bound constraints and show that there is no integral optimal solution of the linear Program~\eqref{eq:ShapleyShubikObjFull}-\eqref{eq:SSdistributionalUB-constraintsFull} under these constraints. In particular, consider a setting with three workers, labelled $w_1$, $w_2$, and $w_3$, and one firm, labelled $f_1$. Further consider the following set of upper  bound constraints: (i) at most one of workers $w_1$ and $w_2$ can be hired, i.e., $x_{w_1, f_1} + x_{w_2, f_1} \leq 1$, (ii) at most one of workers $w_2$ and $w_3$ can be hired, i.e., $x_{w_2, f_1} + x_{w_3, f_1} \leq 1$, and (iii) at most one of workers $w_1$ and $w_3$ can be hired, i.e., $x_{w_1, f_1} + x_{w_3, f_1} \leq 1$, as depicted in Figure~\ref{fig:non-integrality}. Furthermore, suppose that the match values of all the worker-firm pairs are equal, i.e., $\alpha_{w_1,f_1} = \alpha_{w_2,f_1} = \alpha_{w_3,f_1} = 1$. Then, in this setting, the unique optimal solution to the linear Program~\eqref{eq:ShapleyShubikObjFull}-\eqref{eq:SSdistributionalUB-constraintsFull} is $x_{w_1,f_1} = 0.5$, $x_{w_2,f_1} = 0.5$ and $x_{w_3,f_1} = 0.5$ which achieves a total match value of $1.5$, while any feasible integral assignment would result in a total match value of at most one. Thus, there are market instances when the optimal solution of the linear Program~\eqref{eq:ShapleyShubikObjFull}-\eqref{eq:SSdistributionalUB-constraintsFull} may not be integral.

\subsection{Distributional Constraints Alter the Properties of Stable Arrangements} \label{apdx:nonUniqueness}

\subsubsection{Stable Arrangements Beyond LP Duality}

We construct an instance to show that there may be payoff vectors $\u, \v$ for the workers and firms,  respectively, that are not solutions to the dual linear Program~\eqref{eq:DualObjUB}-\eqref{eq:signConstraintsUB} but are still associated with some stable arrangement. To this end, consider a setting of one worker $w$ and one firm $f$, where the firm has an upper bound constraint on hiring at most two workers. Further, suppose that the value of the firm for the worker is given by $c_{w,f} = 1$ and the value of the worker for the firm is given by $a_{w,f} = 0.5$. Then, under the solution of the dual linear Program~\eqref{eq:DualObjUB}-\eqref{eq:signConstraintsUB} the payoff for the firm ($v_{f} = \Bar{\eta}_{\{ w\}} = 0$) is zero by complimentary slackness since the upper bound constraint on hiring is not met with equality. However, one can observe that any salary $s_{w,f}$ such that $-0.5 \leq s_{w,f} \leq 1$ will result in a stable arrangement where the worker and the firm are assigned to each other. That is, there are stable arrangements at which the payoffs of the workers and firms do not correspond to solutions of the dual linear Program~\eqref{eq:DualObjUB}-\eqref{eq:signConstraintsUB}, e.g., at a salary of $s_{w,f} =0.5$, the firm has a positive payoff.

\subsubsection{Structure of Payoffs of Firms and Workers}

We now present an example to show that in the many-to-one matching setting under distributional constraints, the payoffs of firms and workers do not form a lattice structure, as is the case in the one-to-one matching setting of Shapley and Shubik~\cite{shapley1971assignment}. In particular, Shapley and Shubik~\cite{shapley1971assignment} established in the one-to-one matching setting that if the payoffs $\u, \v$ and $\u', \v'$ correspond to two stable arrangements, then the payoffs $\Bar{\u}, \underline{\v}$ and $\underline{\u}, \Bar{\v}$ also correspond to stable arrangements, where $\Bar{u}_w = \max \{u_w, u_w' \}$, $\underline{u}_w = \min \{u_w, u_w' \}$, $\Bar{v}_f = \max \{ v_f, v_f' \}$, and $\underline{v}_f = \min \{ v_f, v_f' \}$. We show that the payoffs $\Bar{\u}, \underline{\v}$ and $\underline{\u}, \Bar{\v}$ may not correspond to any stable arrangements in the many-to-one matching setting with distributional constraints through the following example.


Consider a setting of two workers, labeled $w_1$ and $w_2$, and one firm, labeled $f_1$, where the firm has linear preferences and upper and lower bound constraints on hiring the two workers given by $2 \leq x_{w_1,f_1} + x_{w_2,f_1} \leq 2$. These constraints imply that the only feasible allocation is when the firm hires both workers. Next, suppose that the values of the two workers for the firm are given by $a_{w_1,f_1} = 0$, $a_{w_2,f_1} = 0$, while the value of the firm for the two workers is given as $c_{w_1,f_1} = 2.5$ and $c_{w_2,f_1} = 0.5$. Given these values, the efficient allocation $\Xb^* = (1,1)$ is to match both workers to the firm rather than remaining unmatched. Furthermore, under the salaries $\s = (1, 1)$ and $\s' = (2.5, 0.5)$, we observe that the vector of payoffs are $\u = (1, 1)$, $v_{f_1} = 1$, and $\u' = (2.5, 0.5)$ and $v_{f_1}' = 0$, respectively.

We first show that the arrangements $(\Xb^*, \s)$ and $(\Xb^*, \s')$ are stable and then establish that any arrangement corresponding to the assignment $\Xb^*$ and the payoffs $\Bar{\u}, \underline{\v}$ (or $\underline{\u}, \Bar{\v}$) is not stable.

To see that $(\Xb^*, \s)$ is stable, we first note that the arrangement is feasible since the assignment $\Xb^*$ is feasible as it respects the firm's distributional constraints. Next, this arrangement is certainly individually rational for workers and firms. Finally, it is also clear that there is no other salary profile at which both the workers and firm could be weakly better off with at least one of the two workers or the firm being strictly better off. This is because a strict increase in the salary of any worker would result in a decreased payoff for the firm, and vice versa. Thus, $(\Xb^*, \s)$ is a stable arrangement. Using a very similar line of reasoning it can be shown that $(\Xb^*, \s')$ is also a stable arrangement.

Now, we establish that an arrangement corresponding to the payoffs $\Bar{\u}, \underline{\v}$ is not stable. To see this, observe that $\Bar{u}_{w_1} + \Bar{u}_{w_2} + \underline{v}_{f_1} = 2.5+1+0>3 = \alpha_{w_1,f_1}+\alpha_{w_2,f_1}$, which contradicts stability since the sum of the payoffs must add up to the total match value.

Next, the arrangement corresponding to the payoffs $\underline{\u}, \Bar{\v}$ is not stable. To see this, observe that $\underline{u}_1 + \underline{u}_2 + \Bar{v}_1 = 1+0.5+1 = 2.5<3=\alpha_{w_1,f_1}+\alpha_{w_2,f_1}$, which again contradicts stability since the sum of the payoffs must add up to the total match value.

Thus, we have shown that in the many-to-one matching setting under distributional constraints, the payoffs of firms and workers do not form a lattice structure.

\subsection{Individual Rationality and Feasibility May be at Odds with Lower Bound Constraints} \label{apdx:IRodds}

We construct an instance to show that any feasible assignment that respects the lower bound constraints on hiring for a firm may not correspond to an individually rational outcome for that firm. To this end, consider a setting with one firm $f$ and one worker $w$, where the firm has a lower bound constraint on hiring the worker with a quota of one, i.e., $x_{w,f} \geq 1$. That is, assigning the worker to the firm is the only feasible assignment. Now, if the match value of the worker-firm pair is strictly less than zero, i.e., $\alpha_{w,f}<0$, then it must hold that $u_w + v_f = \alpha_{w,f}<0$. Thus, the only feasible assignment must violate individual rationality for either the firm or the worker. If the resulting outcome is such that it is not individually rational for the worker, then the worker is better off remaining unmatched and thus receiving a payoff of zero. As a result, the only feasible assignment of hiring the worker can only be supported through a negative payoff to the firm.

\end{document}

%% file: tex/preamble.tex
\usepackage{graphicx}
\usepackage{csvsimple}
\usepackage{rotating}
\usepackage[letterpaper, portrait, margin=1in]{geometry}
\usepackage{hyperref}
\hypersetup{
    colorlinks=true,
    linkcolor=blue,
    citecolor=red,
    filecolor=magenta,      
    urlcolor=cyan
    }

\usepackage[nocomma]{optidef}
\usepackage{listings}
\usepackage{comment}
\usepackage{appendix}
\usepackage[ruled,vlined]{algorithm2e}
\usepackage{amsfonts} 
\usepackage{amssymb}
\usepackage{bm}

\usepackage{color} 
\definecolor{mygreen}{RGB}{28,172,0} 
\definecolor{mylilas}{RGB}{170,55,241}
\DeclareFixedFont{\ttb}{T1}{txtt}{bx}{n}{12} 
\DeclareFixedFont{\ttm}{T1}{txtt}{m}{n}{12}  
\usepackage{bbm}
\usepackage{amsmath,amsthm}

\newtheorem{theorem}{Theorem}
\newtheorem{corollary}{Corollary}
\newtheorem{proposition}{Proposition}

\newtheorem{lemma}{Lemma}

\theoremstyle{definition}
\newtheorem{definition}{Definition}
\newtheorem{remark}{Remark}
\newtheorem{example}{Example}

\newenvironment{hproof}{%
  \proof}{\endproof}

\usepackage{color}
\definecolor{deepblue}{rgb}{0,0,0.5}
\definecolor{deepred}{rgb}{0.6,0,0}
\definecolor{deepgreen}{rgb}{0,0.5,0}

\usepackage{listings}

\renewcommand\footnotemark{}

%% file: tex/macros.tex
\def\x{\bm{x}}
\def\v{\bm{v}}
\def\u{\bm{u}}

\def\r{\bm{r}}
\def\q{\bm{q}}

\def\y{\bm{y}}
\def\c{\bm{c}}
\def\b{\bm{b}}

\def\s{\bm{s}}

\def\H{\mathcal{H}}
\def\Xb{\mathbf{X}}

\def\Yb{\mathbf{Y}}
\def\Yc{\mathcal{Y}}

\def\T{\mathcal{T}}
\def\0{\bm{0}}

\def\llambda{\bm{\lambda}}
\def\eeta{\bm{\eta}}
\def\mmu{\bm{\mu}}

\DeclareMathOperator*{\argmax}{arg\,max}